\documentclass[12pt]{article}
\usepackage{a4wide}
\usepackage[dvips]{graphicx,psfrag}
\usepackage{amsfonts}
\usepackage{amstext}
\usepackage{amsmath}
\usepackage{amssymb}
\usepackage{amsthm}
\usepackage[mathscr]{eucal}
\makeatletter
 
 \@addtoreset{equation}{section}
\makeatother
\def\a{\alpha}

\def\Ai{{\rm Ai}}
\def\b{\beta}

\def\e{\epsilon}
\def\E{\mathbb{E}}
\def\g{\gamma}

\def\i{\infty}

\def\K{\mathcal{K}}

\def\L{\mathcal{L}}

\def\o{\omega}

\def\1{\bf{1}}
\def\R{\mathbb{R}}

\def\t{\tau}

\def\k{\kappa}

\def\ss#1{\fontfamily{cmss}\fontseries{m}\fontshape{n}\selectfont #1}
\begin{document}
\title{Replica approach to the KPZ equation with\\
half Brownian motion initial condition}

\author{Takashi Imamura
\footnote { Research Center for Advanced Science and Technology
, The University of Tokyo, E-mail: imamura@jamology.rcast.u-tokyo.ac.jp}
, Tomohiro Sasamoto
\footnote { Department of mathematics and informatics, 
Chiba University, E-mail: sasamoto@math.s.chiba-u.ac.jp}}

\maketitle

\begin{abstract}
We consider the one-dimensional Kardar-Parisi-Zhang (KPZ) equation 
with half Brownian motion initial condition,
studied previously through the weakly asymmetric simple exclusion 
process. We employ the replica Bethe 
ansatz and show that the generating function of the exponential 
moments of the height is expressed as a Fredholm determinant. 
From this the height distribution and its asymptotics are studied. 
Furthermore using the replica method we also discuss the multi-point 
height distribution. We find that some nice properties of the deformed 
Airy functions play an important role in the analysis. 
\end{abstract}
\section{Introduction}
Surface growth phenomena appear widely in nature 
and have attracted much attention in non-equilibrium physics.
In 1986, Kardar, Parisi and Zhang proposed a stochastic 
differential equation which describes surface growth with
local interaction~\cite{KPZ1986}. For one-dimensional case,
the KPZ equation is given by
\begin{equation}
\frac{\partial h(x,t)}{\partial t}=\frac{\lambda}{2}
\left(\frac{\partial h(x,t)}{\partial x}\right)^2+
\nu\frac{\partial^2 h(x,t)}{\partial x^2}+\sqrt{D}\eta(x,t).
\label{KPZ}
\end{equation}
Here $h(x,t)$ represents the height of the surface at position 
$x\in \R$ and time $t\ge 0$. The first term represents
the effect of nonlinearity and the second one describes the 
smoothing effect of the surface.
$\eta (x,t)$ represents randomness described by
the Gaussian white noise with covariance,
\begin{equation}
\langle \eta(x,t)\eta(x',t')\rangle=\delta(x-x')\delta(t-t').
\label{random}
\end{equation}
The parameters $\lambda,~\nu,~D$ determine their strengths. 
For the one-dimensional KPZ equation~\eqref{KPZ}, 
it has been shown by the dynamical renormalization group that
the height fluctuation scales as $O(t^{1/3})$ as
$t$ goes to infinity~\cite{KPZ1986}. The exponent $1/3$ coincides with the
ones found in Monte Carlo simulations of many stochastic models of 
surface growth. The KPZ equation is accepted as a prototypical equation 
describing the universality class called the KPZ universality class.

Our understanding of the KPZ universality class has deepened
in the last decade. Not merely the exponent but the height distribution 
functions
have been computed based on intriguing connections with the random 
matrix theory~\cite{Jo2000}.
For the totally asymmetric simple exclusion process(TASEP), which is
an exactly solvable model in the KPZ universality class,
the current distribution function for the step initial condition
has been obtained exactly and 
it was found that in the long time limit, it converges to 
the GUE (Gaussian Unitary Ensemble) Tracy-Widom distribution 
function~\cite{TW1994}, the largest eigenvalue distribution 
in random hermitian 
matrix~\cite{Jo2000}. 

An interesting finding from the studies of distribution functions is that 
they detect the difference of initial conditions, 
which the scaling exponent of the height cannot.
For example, in the surface growth model called polynuclear growth(PNG)
model, it has been recognized that the GUE Tracy-Widom 
distribution describes the height distribution in the droplet 
growth~\cite{PS2000a}
while in the flat initial condition, it is described by the GOE
(Gaussian Orthogonal Ensemble) Tracy-Widom distribution~\cite{PS2000b}.
Furthermore, generalizations of these results to 
multi-point distribution function have also 
been studied and the universal processes, 
the Airy$_2$~\cite{PS2002b} 
and Airy$_1$~\cite{Sa2005,BFPS2007} processes
have been obtained. For recent progresses of this topic, 
see~\cite{S2007r,KK2010,F2010}.

Since 2010, the studies on the KPZ equation and KPZ universality 
class have entered a new stage~\cite{SS2010d}. 
First as an experimental progress, high-accuracy measurements 
for the exponents and the height distribution in the KPZ growth 
problem using turbulent liquid crystal have been performed and 
the GUE Tracy-Widom distribution function was observed as well as
the critical exponents~\cite{TS2010}. 
Second, we have begun to understand exactly height distribution
function of the KPZ equation itself. 
In \cite{SS2010a,SS2010b, SS2010c,ACQ2010} the height distribution at 
a single position was computed for the narrow-wedge initial condition,
\begin{equation}
\frac{\lambda}{2\nu}h(x,t=0)=-\frac{|x|}{\delta}, 
~\delta\rightarrow 0,
\label{narrow}
\end{equation}
from which the surface grows to a parabolic shape. 
In the long-time limit, the fluctuation around the macroscopic shape 
is described by the GUE Tracy-Widom distribution.
The analysis of \cite{SS2010a,SS2010b, SS2010c,ACQ2010} 
is based on recent progress on the current distribution of the 
(partially) ASEP~\cite{TW2008,TW2009} and 
a fact that, in the weak asymmetry limit, the stochastic time 
evolution of the current of the ASEP can be mapped to that 
of the height described by the KPZ equation~\cite{BG1997}. 
For more recent developments, see~\cite{FF2011p,QR2011p,O2009p,OW2011p,GJ2010p1,GJ2010p2}.

More direct approach without relying on the results on the ASEP
has also been developed recently \cite{D2010,Dotsenko2010,CLDR2010}.
The idea was first proposed in~\cite{K1987}, in which  
the author used the interesting relation that the exponential 
moment of the height is represented as the dynamics of the one-dimensional 
$\delta$-function Bose gas with attractive interaction,
which is solved by the Bethe ansatz~\cite{LL1963,M1964}.
In the original work~\cite{K1987}, only the ground state contribution
was considered and the dynamical exponent $1/3$ was
obtained.
In~\cite{Dotsenko2010,CLDR2010}, the authors succeeded in
taking into account the whole contribution of the eigenstates
and obtained the Fredholm determinant representation of
the generating function of the exponential moment for the narrow
wedge initial condition~\eqref{narrow}. 
More recently a multi-point generating function 
is also discussed~\cite{PS2010p, PS2010p2}. 

The replica method is quite attractive since we expect to 
be able to calculate various quantities for the one-dimensional 
KPZ equation directly and easily. In particular, we 
expect it would be a powerful tool 
to understand the universality in the KPZ equation. 
There are two recent progresses in this direction:
In~\cite{CQ2011p}, the authors discussed a  
process characterizing  the renormalization fixed point of the 
KPZ universality class and the Fredholm determinant expression
of its transition probability has been obtained. 
In~\cite{CLD2011p},
the authors have obtained the exact height distribution
for the flat initial condition and clarified its convergence
to the GOE Tracy-Widom distribution in the long time limit. 

In this paper, we pursue the potential
of the replica method by applying it to
the half Brownian-motion initial condition depicted in 
Fig.~1. This is one of the typical spatially extended 
initial conditions and is written in terms of the single 
valued function $h(x,t)$ as
\begin{align}
\frac{\lambda}{2\nu}h(x,t=0)&=
\begin{cases}
x/\delta, 
~\delta\rightarrow0, & x<0,\\
\a B(x), & x\ge 0.
\end{cases}
\label{stepBM}
\end{align}
Here $B(x)$ represents the one-dimensional standard
Brownian motion with $B(0)=0$ and $\a=(2\nu)^{-3/2}\lambda D^{1/2}$. 
For this initial condition, the macroscopic shape 
expected by solving~\eqref{KPZ} without the noise term is 
\begin{align}
h(x,t)\sim
\begin{cases}
-x^2/2\lambda t,& x \le 0, \\
0, & 0<x.
\end{cases}
\label{macro}
\end{align}
The main interest in this paper is 
the fluctuation around 
this macroscopic shape. 
We expect that a one-dimensional Brownian motion describes the fluctuation 
in the positive region ($x>0$) since it is known to be a stationary measure of
the KPZ equation. On the other hand, in the negative region ($x<0$) where
the parabolic growth is observed the situation is the same as the narrow 
wedge case. Thus the fluctuation property around the origin shows a crossover
behavior between the two typical growths.

This initial condition has already been considered in \cite{CQ2010p}. 
There the analysis is based on the result for the ASEP with step Bernoulli 
initial condition~\cite{TW2009b} and naturally leads to an expression 
for the height distribution in the form of contour integral. We take 
the replica method and mainly treat the generating function of the 
exponential moment. Accordingly our formula is somewhat different 
from the one in \cite{CQ2010p} and is expressed in terms only of 
real quantities. Furthermore the replica approach allows us to discuss
the multi-point distribution function with the help of the factorization
assumption by \cite{PS2010p,PS2010p2} and properties of the deformed 
Airy functions discussed in Appendix.

\begin{figure}
\begin{picture}(200,120)
\put(100,0){\includegraphics[width=200pt]{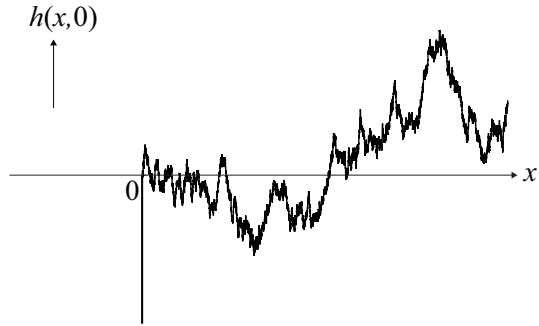}}
\end{picture}
\caption{Half Brownian motion initial condition
}
\end{figure}

This paper is arranged as follows. In the next section, 
we state our main results. 
Briefly explaining the relation between the KPZ equation
and the $\delta$-Bose gas in Sec.~3, we give a derivation
of our main result, the Fredholm determinant expression of the 
generating function (Theorem 1 in Sec.~2) in Sec.~4. 
In Sec.~5, we discuss
another expression of Theorem 1 stated as Proposition 2 in Sec.~2, 
which is useful for the compact expression of the height distribution 
function described by Theorem 3 in Sec.~2. 
In Sec.~6, we discuss the multi-point height distribution.
In Sec.~7, we consider the interpretation of our result as 
free energy distribution of a directed polymer in random media.
Concluding remarks are 
given in the last section.

\section{Model and main results}
\subsection{The generating function}
The KPZ equation (\ref{KPZ}) is in fact ill-defined as it is.  
As time goes on the height profile approaches the stationary 
one described by the Brownian motion, for which the non-linear 
term in the KPZ equation is not well-defined. 
A proper prescription is proposed in~\cite{BG1997}, 
which we follow here. 
In this scheme, one {\it defines} the height of the KPZ equation by
\begin{align}
h_{\nu,\lambda,D}(x,t)=\frac{2\nu}{\lambda}
\log\left(Z_{\nu,\lambda,D}(x,t)\right),
\label{ColeHopf0}
\end{align}
using the solution $Z_{\nu,\lambda,D}(x,t)$ of the stochastic 
heat equation,
\begin{equation}
\frac{\partial Z_{\nu,\lambda,D}(x,t)}{\partial t}=
\nu\frac{\partial^2 Z_{\nu,\lambda,D}(x,t)}
{\partial x^2}+\frac{\lambda\sqrt{D}}{2\nu}\eta(x,t)Z_{\nu,\lambda,D}
(x,t),
\label{SHE}
\end{equation}
which is a well-defined stochastic partial differential equation of 
It{\^o}-type. This is called the Cole-Hopf solution of the KPZ 
equation because the equation (\ref{SHE}) is related to the KPZ 
equation through the (inverse of) the Cole-Hopf
transformation (\ref{ColeHopf0}). We will give more explanations
in subsection 3.1 below. Hereafter we investigate the properties of these
regularized quantities $h_{\nu,\lambda,D}(x,t)$ and 
$Z_{\nu,\lambda,D}(x,t)$
with the initial conditions~\eqref{stepBM}. 
In the following we denote by $\langle\cdots\rangle$ the average over 
both $\eta(x,t)$ in the KPZ equation~\eqref{KPZ} and 
$B(x)$ in the initial condition~\eqref{stepBM}.

We are interested in the distribution of the height 
$h_{\nu,\lambda,D}(x,t)$. 
It is well established, known as the KPZ scaling, that the fluctuation of 
the height scales like $O(t^{1/3})$ and nontrivial correlations are 
seen in the $x$ direction with scale $O(t^{2/3})$ when $t$ is large.  
Let us define a parameter $\gamma_t$ which scales as $O(t^{1/3})$ and 
a rescaled space coordinate $X$ by 
\begin{align}
\gamma_t=\left(\a^4 \nu t\right)^{\frac13},~~
x=\frac{2\gamma_t^2X}{\a^2},
\label{generalgammaX}
\end{align}
with $\a$ given below~\eqref{stepBM}.  
We introduce the scaled height $\tilde{h}_t(X)$ by 
\begin{equation}
\frac{\lambda}{2\nu}h_{\nu,\lambda,D} 
\left( \frac{2\gamma_t^2X}{\a^2},t \right)
=
-\frac{\gamma_t^3}{12}-\gamma_tX^2 + \gamma_t \tilde{h}_t(X).
\label{scaled}
\end{equation}
Here the second term corresponds to the macroscopic shape in
(\ref{macro}); it is an interesting aspect of the KPZ equation 
that one has also to take into account the first term to focus 
on the height fluctuations.

To study the distribution, it is often useful to consider
the generating function of the moments
$\langle Z_{\nu,\lambda,D}(x,t)^N\rangle,$~
$N=0,1,2,\cdots$. We define $G_{\gamma_t}(s;X)$ as 
\begin{equation}
G_{\gamma_t}(s;X)
=
\sum_{N=0}^{\infty}\frac{\left(-e^{-\gamma_t s}\right)^N}{N!}
\left\langle 
{Z}_{\nu,\lambda,D}^N \left( \frac{2\gamma_t^2X}{\a^2},t \right)
\right\rangle  
e^{N\frac{\gamma_t^3}{12} +N\gamma_t X^2} 
=
\langle e^{-e^{\gamma_t(\tilde{h}_t(X)-s)}}\rangle.
\label{gr}
\end{equation}
By using the replica method, one can express the moment 
$\langle Z_{\nu,\lambda,D}(x,t)^N\rangle$ of our problem in  
the language of the $\delta$-function Bose gas, which is a 
well-known exactly solvable model. One can further perform the 
summation over $N$ to obtain the Fredholm determinant representation
of the generating function.

\vspace{5mm}
\noindent
{\bf Theorem 1}
{\it $G_{\gamma_t}(s;X)$ is expressed as the Fredholm determinant
with the kernel acting on $L^2(\R)$,
\begin{align}
~G_{\gamma_t}(s;X)
&=\det\left(1-P_0 K_XP_0\right).
\label{KPZfred1}
\end{align}
Here $P_s$ represents the projection onto $(s,\infty)$ and
the kernel of $K_X$ is given by 
\begin{equation}
K_X(\o_j,\o_k)=
\sum_{n=1}^{\infty}(-1)^{n-1}
\int_{\R-ic_n}\frac{dq}{\pi}
e^{-n(\o_j+\o_k)-2iq(\o_j-\o_k)-\gamma_t^3nq^2+\frac{\gamma_t^3}{12}n^3
-\gamma_t ns}
\frac{\Gamma\left(iq-\frac{X}{\gamma_t}-\frac{n}{2}\right)}
{\Gamma\left(iq-\frac{X}{\gamma_t}+\frac{n}{2}\right)},
\label{th3kernel1}
\end{equation}
where $\Gamma (x)$ is the gamma function and 
$c_n$ satisfies $c_n>X/\gamma_t+n/2$. }

\vspace{5mm}
\noindent
Here for an operator on $L^2(\mathbb{R})$ with kernel $K(x,y)$, 
the Fredholm determinant is defined by 
\begin{equation}
\det\left(1- K \right)
=
\sum_{M=0}^\infty\frac{(-1)^M}{M!}
\int_{-\infty}^\infty dx_1\cdots \int_{-\infty}^{\infty}dx_M
\det\left(K(x_j,x_k)
\right)_{j,k=1}^M 
\end{equation}
where the right hand side is assumed to converge.

A derivation of this result will be given in Sec.~4 after the 
explanation of the relation of $N$th moment of $Z_{\nu,\lambda,D}(x,t)$ 
with $N$ particle dynamics of the $\delta$-function Bose gas in Sec.~3.
For the narrow wedge case~\eqref{narrow} where 
$Z_{\nu,\lambda,D}(x,0)=\delta(x)$, the corresponding generating function
has been obtained in the same form 
as~\eqref{KPZfred1}~\cite{Dotsenko2010, CLDR2010}. 
But the kernel $K_X(\o_j,\o_k)$ is replaced by 
\begin{equation}
K_{\text{nw}}(\o_j,\o_k)=
\sum_{n=1}^{\infty}(-1)^{n-1}
\int_{-\infty}^{\infty}\frac{dq}{\pi}
e^{-n(\o_j+\o_k)-2iq(\o_j-\o_k)-\gamma_t^3nq^2+\frac{\gamma_t^3}{12}n^3
-\gamma_t ns},
\label{th3kernelnw}
\end{equation}
which is independent of the position $X$.

One can replace the kernel by the one written in the form of 
products of two deformed Airy functions.   

\vspace{3mm}
\noindent
{\bf Proposition 2}
{\it In~\eqref{KPZfred1}, the kernel $P_0K_XP_0$ can be replaced by
$P_0\bar{K}_XP_0$ where
\begin{align}
\bar{K}_X(\xi_j,\xi_k)
=
\int_{-\infty}^{\infty} dy \Ai^\Gamma\left(\xi_j+y,\frac{1}{\gamma_t},-\frac{X}{\gamma_t}\right)
\Ai_{\Gamma}\left(\xi_k+y,\frac{1}{\gamma_t},-\frac{X}{\gamma_t}\right)
\frac{e^{\gamma_t y}}{e^{\gamma_t y}+e^{\gamma_t s}}.
\label{alternative1}
\end{align}
Here $\Ai^{\Gamma}(a,b,c),~\Ai_{\Gamma}(a,b,c)$ are defined by
\begin{align}
&\Ai^\Gamma (a,b,c)=\frac{1}{2\pi}
\int_{\Gamma_{i\frac{c}{b}}} dz 
e^{iza+i\frac{z^3}{3}}{\Gamma\left(ibz+c\right)},
\label{AiG1}\\
&\Ai_\Gamma (a,b,c)=\frac{1}{2\pi}
\int_{-\infty}^{\infty} dz 
e^{iza+i\frac{z^3}{3}}\frac{1}{\Gamma\left(-ibz+c\right)}.
\label{AiG2}
\end{align}
In~\eqref{AiG1},
$\Gamma_{z_p}$ represents the contour  from 
$-\infty$ to $\infty$ and, along the way, 
passing below the pole $z_p=ic/b$.
}

\vspace{3mm}

\noindent
The functions $\Ai^{\Gamma}(a,b,c),~\Ai_{\Gamma}(a,b,c)$ 
have appeared in~\cite{CQ2010p}. Note that these become 
the ordinary Airy function if the gamma function factors in the 
integrand are eliminated in~\eqref{AiG1} and~\eqref{AiG2}. 
For the narrow wedge case, the corresponding kernel is the one where
$\Ai^{\Gamma}(x,1/\gamma_t,-X/\gamma_t)$ and
$~\Ai_{\Gamma}(x,1/\gamma_t,-X/\gamma_t)$ are replaced by
the Airy function. To get the kernel in proposition 2 from that 
in Theorem 1, one has to find some generalizations of formulas 
utilized in \cite{Dotsenko2010,CLDR2010}. 
They and a derivation of (\ref{alternative1}) will be given in Sec.~5.  
Some properties of the deformed Airy functions 
are summarized in Appendix.

\subsection{The height distribution function}
The information of all the moments is enough to extract that of  
the probability distribution function. 
Applying the discussions in~\cite{CLDR2010,PS2010p}, we can 
find a formula of the height distribution from the generating 
function $G_{\gamma_t}(s,X)$. Let $F_{\gamma_t}(s;X)$ be 
\begin{align}
F_{\gamma_t}(s;X)=\text{Prob}\left(\frac{\lambda}{2\nu}
h(x,t)+\frac{\gamma_t^3}{12}+\gamma_tX^2 \le \gamma_t s \right)
=\text{Prob}(\tilde{h}_t \leq s).
\label{hdist}
\end{align}
By using the Fredholm determinant~\eqref{KPZfred1}, 
$F_{\gamma_t}(s;X)$ can be expressed as 
follows~\cite{CLDR2010,PS2010p}.
\begin{align}
F_{\gamma_t}(s;X)
=1-\int_{-\infty}^{\infty}
du e^{-e^{\gamma_t(s-u)}}
g_{\gamma_t}(u;X).
\end{align}
Here
\begin{align}
g_{\gamma_t}(u;X)=\frac{1}{2\pi i}
\left(
\det(1-P_0K^+_XP_0)-\det(1-P_0K^-_XP_0)
\right)
\label{gexpression}
\end{align}
where $K^{\pm}_X(x,y)$ is the kernel~\eqref{th3kernel1} or~\eqref{alternative1}
in which the term $e^{-s}$ is replaced by $-e^{u}\pm i\e$
with $\e>0$ being infinitesimal.

Using~\eqref{alternative1}, and the relation
$1/(x\pm i\e)= \mathcal{P}(1/x)\mp i\pi\delta(x)$, where
$\mathcal{P}$ denotes the Cauchy principal value,
we can easily find
$K^{\pm}_X$ in~\eqref{gexpression} is represented as
\begin{align}
K^{\pm}_X(\xi_j,\xi_k)
&=\mathcal{P}\int_{-\infty}^{\infty} dy \Ai^\Gamma\left(\xi_j+y,
\frac{1}{\gamma_t},-\frac{X}{\gamma_t}\right)
\Ai_{\Gamma}\left(\xi_k+y,\frac{1}{\gamma_t},-\frac{X}{\gamma_t}\right)
\left(\frac{1}{1-e^{\gamma_t(u-y)}}\right)\notag\\
&~~~~
\mp i\pi \Ai^\Gamma\left(\xi_j+u,\frac{1}{\gamma_t},-\frac{X}{\gamma_t}\right)
         \Ai_{\Gamma}\left(\xi_k+u,\frac{1}{\gamma_t},
-\frac{X}{\gamma_t}\right).
\end{align}
Substituting this expression to~\eqref{gexpression} and using
basic properties of determinant, we eventually arrive
at the expression in terms of the Fredholm determinant. 

\vspace{5mm}
\noindent
{\bf Theorem 3}
{\it
\begin{align}
F_{\gamma_t}(s;X)&
=
1-\int_{-\infty}^{\infty}
du e^{-e^{\gamma_t(s-u)}}
g_{\gamma_t}(u;X).
\label{Th3}
\end{align}
Here $g_{\gamma_t}(u;X)$ is expressed as a difference of two
Fredholm determinants, 
\begin{align}
g_{\gamma_t}(u;X)=\det\left(1-P_u(B^{\Gamma}_{\gamma_t}-P^{\Gamma}_{\Ai})P_u\right)
-\det\left(1-P_uB^\Gamma_{\gamma_t}P_u\right),
\end{align}
where %$P_u$ is a projection operator on $(u,\infty)$, and
\begin{align}
B^{\Gamma}_{\gamma_t}(\xi_1,\xi_2)
&=\int_0^{\infty}dy
\Ai^\Gamma\left(\xi_1+y,
\frac{1}{\gamma_t},-\frac{X}{\gamma_t}\right)
\Ai_\Gamma\left(\xi_2+y,
\frac{1}{\gamma_t},-\frac{X}{\gamma_t}\right)
\notag\\
&~~+\int_{0}^{\infty}dy\frac{1}{e^{\gamma_t y}-1}
\left(
\Ai^\Gamma\left(\xi_1+y,
\frac{1}{\gamma_t},-\frac{X}{\gamma_t}\right)
\Ai_\Gamma\left(\xi_2+y,
\frac{1}{\gamma_t},-\frac{X}{\gamma_t}\right)
\right.\notag\\
&\hspace{3.2cm}\left.-
\Ai^\Gamma\left(\xi_1-y,
\frac{1}{\gamma_t},-\frac{X}{\gamma_t}\right)
\Ai_\Gamma\left(\xi_2-y,
\frac{1}{\gamma_t},-\frac{X}{\gamma_t}\right)\right)
,
\label{BGamma}\\
P^{\Gamma}_{\Ai}(\xi_1,\xi_2)&=
\Ai^\Gamma\left(\xi_1,
\frac{1}{\gamma_t},-\frac{X}{\gamma_t}\right)
\Ai_\Gamma\left(\xi_2,
\frac{1}{\gamma_t},-\frac{X}{\gamma_t}\right)
.
\label{PGamma}
\end{align}
}

\vspace{3mm}

\noindent
In~\cite{CQ2010p}, another expression using a contour integral
was obtained whereas ours~\eqref{Th3} are represented as 
the convolution with the Gumbel distribution. 
For numerical analysis, this form would be convenient.
Actually, from this expression, we can readily draw
the picture of the probability density function $d F_{\gamma_t}(s;X)/ds$ 
as in  Fig.~2. In this figure, the graphs are drawn approximating 
the Fredholm determinant by a finite dimensional matrix determinant 
using a simple discretization. For more precise estimation, the method
in~\cite{Bo2010,Bo20102} 
is available and actually it was applied to the narrow wedge
case~\cite{PS2011p}.

\begin{figure}
\begin{picture}(200,150)
\put(100,0){\includegraphics[width=300pt]{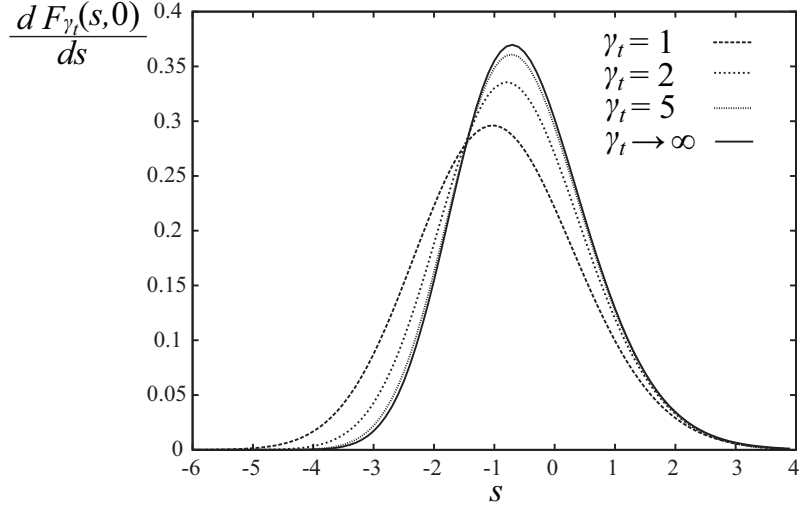}}
\end{picture}
\caption{The probability density functions at the origin $X=0$.
 The three cases for finite $t$'s ($\g_t=1,2,5$) are depicted as the
 dashed lines while the solid line corresponds to the 
$\gamma_t \to \infty$ limit (see~\eqref{Cor4}). 
}
\end{figure}

The distribution function~\eqref{Th3} has a similar form as
the narrow wedge case obtained 
in~\cite{SS2010a,SS2010b,SS2010c,ACQ2010}:
If we replace the functions $\Ai^\Gamma(x)$ and
$\Ai_{\Gamma}(x)$ by the ordinary Airy function in~\eqref{BGamma} 
and~\eqref{PGamma}, the distribution
function~\eqref{Th3} becomes the one for the narrow wedge initial 
condition. 

We also consider the long-time 
limit ($t\rightarrow\infty$) of the distribution function.
Let us remember a basic fact that the probability distribution function 
is in general written as the expectation,
$\text{Prob}(X\leq s) = \E(\Theta(s-X))$,
where $\Theta(y)$ is the step function, 
$\Theta(y):=1~(y\ge 0),~0~(y<0)$. 
Noticing $\displaystyle\lim_{a\to\infty} \exp[-e^{-ax}] = \Theta(x)$,
we take the $t\to\infty$ limit in (\ref{gr})
and see
\begin{align}
\lim_{\gamma_{t}\rightarrow\infty}\text{Prob}(\tilde{h}_t\le s)
=\lim_{\gamma_{t}\rightarrow\infty}G_{\gamma_t}(s;X).
\label{lg}
\end{align}
Combining this relation with Theorem 1 and Proposition 2, one obtains
\begin{align}
\lim_{\gamma_t\rightarrow\i}\text{Prob}(\tilde{h}_t\le s)&
=\det\left(1-P_s\K_X P_s\right),
\label{Cor4}
\end{align}
the kernel is given by 
\begin{align}
\K_X(\xi_j,\xi_k)&:=\lim_{\gamma_{t}\rightarrow\infty} 
\bar{K}_X(\xi_j-s,\xi_k-s)\notag\\
&=
\int_{0}^{\infty}dy 
\Ai(\xi_j+y)\Ai(\xi_k+y)\notag\\
&\hspace{3cm}+\Ai(\xi_k)\left(e^{-\frac{X^3}{3}+X \xi_j}-
\int_{0}^{\infty}dye^{-Xy} 
\Ai(\xi_j+y)\right).
\end{align}
This distribution function was obtained in the study on 
the PNG model with an external source~\cite{IS2004,IS2005} 
and the TASEP with step Bernoulli initial condition~\cite{CFP2010}.
At $X=0$, another expression using the solution to the Painlev{\'e} 
differential equation was given in~\cite{BR2000}. 
In Fig. 2, we illustrate the picture of the probability 
density function at the origin $X=0$ as a solid curve.
The derivation of~\eqref{Cor4} will be given in Sec. 6 
including the discussion of the multi-point height distribution.

\subsection{Multi-point distribution function}
The replica approach allows us to discuss multi-point height distribution
function. We introduce the $n$-point generating function 
\begin{align}
G_{\gamma_t}(\{s\}_n,\{X\}_n)
=
\left\langle e^{-\sum_{j=1}^ne^{\gamma_t(\tilde{h}_t(X_j)-s_j)}}
\right\rangle,
\label{multdist}
\end{align}
where we abbreviated $s_1,\cdots,s_n$ and 
$X_1,\cdots,X_n$ as $\{s\}_n$ and $\{X\}_n$ respectively and
we set $X_1<X_2<\cdots<X_n$.

In~\cite{PS2010p,PS2010p2}, the authors proposed  
the ``factorization assumption'' for the $n$-point generating function
for the narrow wedge initial condition. In this paper, we show that if
we employ the same approximation to the case of half Brownian motion 
initial condition, we obtain the following result:
\begin{align}
G_{\gamma_t}^{\sharp}(\{s\}_n,\{X\}_n)
&=\det\left(1-Q\right), 
\label{apge}
\end{align}
and the kernel $Q(x,y)$ is given by
\begin{align}
&Q(u_1,u_{n+1})=
\int_{-\infty}^{\infty}du_2\cdots du_n
\langle u_1|e^{(X_1-X_2)H}
|u_2\rangle\cdots \langle u_n|e^{(X_n-X_1)H}L_1|u_{n+1}
\rangle \Phi(\{u-s\}_n),
\label{qexpr}
\end{align}
where $H$ is the Airy Hamiltonian $H=-\frac{\partial^2}{\partial u^2}+u$,
and
\begin{align}
&\Phi(\{x\}_n)=\frac{\sum_{j=1}^ne^{-\gamma_t x_j}}
{1+\sum_{j=1}^ne^{-\gamma_t x_j}},
\label{Phix}\\
&
L_j(x,y)=\int_{0}^{\infty}dw 
\Ai_{\Gamma}
\left(w+x,\frac{1}{\gamma_t},-\frac{X_j}{\gamma_t}\right)
\Ai^{\Gamma}
\left(w+y,\frac{1}{\gamma_t},-\frac{X_j}{\gamma_t}\right),
\label{Lj}
\end{align}
with $j\geq 1$. 
In~\eqref{apge}, we put the symbol $\sharp$ in order to
represent explicitly the fact that this is an expression after 
using the factorization approximation. 
The factor $\langle x|e^{(X_j-X_k)H}|y\rangle$ in~\eqref{qexpr}
can be expressed as
\begin{align}
\langle x|e^{(X_j-X_k)H}|y\rangle
=\int_{-\infty}^{\infty}dze^{-tz}\Ai_{\Gamma}\left(x+z,
\frac{1}{\gamma_t},-\frac{X_j}{\gamma_t}\right)
\Ai^{\Gamma}\left(y+z,
\frac{1}{\gamma_t},-\frac{X_k}{\gamma_t}
\right),
\label{airyairygg}
\end{align}
see~\eqref{drprop}.
For the one-point case $n=1$, however, the method in~\cite{PS2010p,PS2010p2}
is exact. Actually, when $n=1$, we find the Fredholm determinant is equivalent
to the kernel~\eqref{alternative1}:
In the one-point case, the kernel~\eqref{qexpr} becomes
\begin{align}
Q(u_1,u_2)=L_1(u_1,u_2)\Phi({u_1-s_1}).
\end{align}
We divide it into $Q=Q_1Q_2$, where
\begin{align}
&Q_1(u,w)=\Phi(u-s)\Ai^{\Gamma}\left(w+u,\frac{1}{\gamma_t},-\frac{X_1}
{\gamma_t}\right)\chi_0(w),\\
&Q_2(w,u)=\chi_0(w)\Ai_{\Gamma}\left(w+u,\frac{1}{\gamma_t},-\frac{X_1}
{\gamma_t}\right),
\label{divQ}
\end{align}
where $\chi_s(\xi)=1~(s\le\xi < \infty), 0~(-\infty < \xi < s)$
and notice the property of the Fredholm determinant 
$\det(1-Q_1Q_2)=\det(1-Q_2Q_1)$. 
We easily find that $Q_2Q_1(\xi_1,\xi_2)
=\chi_0(\xi_2)\bar{K}_X(\xi_2,\xi_1)\chi_0(\xi_2)$, where $\bar{K}_X(\xi_2,\xi_1)$
is given in~\eqref{alternative1}. 

Because an approximation is involved in the derivation, 
the expression (\ref{apge}) is most likely not exact when $n\geq 2$. 
A validity of employing this approximation is that, for the
narrow wedge case, the generating function becomes the multi-point
distribution function for the Airy$_2$ process in the long-time limit
which is expected from the universality. In this paper we will 
see that the situation is similar for the case of 
the half Brownian motion initial condition. 
In the long-time limit, the multi-point generating function is 
equivalent to the multi-point height distribution function, 
\begin{align}
\lim_{t\rightarrow\infty}G_{\gamma_t}(\{s\}_n,\{X\}_n)
=\lim_{t\rightarrow\infty}
\text{Prob}\left(\tilde{h}_t(X_j)
\le s_j,~j=1,\cdots, n\right).
\label{ltl}
\end{align}
From (\ref{apge}), we find 
\begin{align}
&~\lim_{t\rightarrow\infty}G^\sharp_{\gamma_t}
\left(\{s\}_n,\{X\}_n\right)=\det\left(1-P_s\K_{12}P_s\right)\notag\\
&:=\sum_{M=0}^\infty\frac{(-1)^M}{M!}
\prod_{i=1}^M\left(\sum_{n_i=1}^n\int_{-\infty}^\infty d\xi_i
\chi_{s_{n_i}}(\xi_i)\right)
\det\left(\K_{12}(X_{n_j},\xi_j;X_{n_k},\xi_k)\right)_{j,k=1}^M
\label{ltlsh}
\end{align}
where $\chi_s(\xi)$ is defined below~\eqref{divQ} and
\begin{align}
&\K_{12}(X_1,\xi_1;X_2,\xi_2)=\K_2(X_1,\xi_1;X_2,\xi_2)\notag\\
&\quad +
\Ai(\xi_2)
\left(e^{-\frac{X_1^3}{3}+X_1\xi_1}-
\int_{0}^{\infty}dye^{-X_1 y} 
\Ai(\xi_2+y)\right) ,\\
&\K_2(X_j,\xi_j;X_k,\xi_k)=
\begin{cases}
\int_{0}^{\infty}dy e^{-w(X_j-X_k)}
\Ai(\xi_j+w)\Ai(\xi_k+w),& j\ge k,\\
-\int_{-\infty}^{0}dy e^{-w(X_j-X_k)}
\Ai(\xi_j+w)\Ai(\xi_k+w),& j<k.
\end{cases}
\label{extendedAiry}
\end{align}
The Fredholm determinant with this kernel has appeared in the PNG model and 
TASEP~\cite{IS2004,IS2005,CFP2010}. 
Note that for the one-point case $n=1$, it reduces to~\eqref{Cor4}.

As pointed out in the Conjecture 9 in~\cite{CQ2010p},
it is plausible that in the long-time limit, the 
Fredholm determinant $\det\left(1-P_s\K_{12}P_s\right)$ 
describes the multi-point distribution considering the exact 
result of the one-point case. 
Thus~\eqref{ltlsh} strongly suggests that the factorization 
approximation becomes exact in the long time limit in both narrow wedge and 
half Brownian motion initial conditions. In Sec. 6, we briefly discuss 
the derivation of these two equations~\eqref{apge} and~\eqref{ltlsh}.

\section{KPZ equation and $\delta$-Bose gas}
\subsection{Feynman-Kac formula}
In subsection 2.1, we mentioned that the KPZ equation (\ref{KPZ})
is not well-defined as it is, and defined its solution by
~\eqref{ColeHopf0} and~\eqref{SHE}. Here we explain this using a
limit of a modified version of the KPZ equation,
\begin{align}
\frac{\partial h_{\kappa}(x,t)}{\partial t}=\frac{\lambda}{2}
\left(\frac{\partial h_{\kappa}(x,t)}{\partial x}\right)^2+
\nu\frac{\partial^2 h_{\kappa}(x,t)}{\partial x^2}+\sqrt{D}\eta_{\kappa}(x,t)
-\frac{1}{2}\left(\frac{\lambda\sqrt{D}}{2\nu}
\right)^2C_\kappa(0).
\label{rKPZ}
\end{align}
This looks almost the same as the original (\ref{KPZ}), but there are
two differences. First $\eta_\kappa(x,t)$ is still a Gaussian noise but
its covariance in $x$-direction is now smeared as
\begin{align}
\langle \eta_\kappa(x,t)\eta_\kappa(x',t')\rangle=C_{\kappa}(x-x')\delta(t-t'),
\label{srandom}
\end{align}
where $C_{\kappa}(x)$ is written as $C_{\kappa}(x)=\kappa C(\kappa x)$ 
and $C(x)$ is a smooth, even and positive function such that 
$\int_{-\infty}^{\infty} C(x)=1$. 
Second, the constant velocity term $-\left({\lambda\sqrt{D}}/{2\nu}
\right)^2C_\kappa(0)/2$ is added. 
Note that the above properties of $C_{\kappa}$ imply 
$\lim_{\kappa\rightarrow\infty}C_\kappa(x)=\delta(x)$
so that the noise $\eta_{\kappa}$ tends to the white noise in the 
original KPZ equation. An apparent problem is that the additional 
term in~\eqref{rKPZ} diverges as $\delta(0)$ but
in fact this term is necessary for considering a meaningful limit of 
the equation (\ref{rKPZ}). %$\kappa\rightarrow\infty$. 
To see this let us apply the Cole-Hopf transformation, 
\begin{equation}
Z_{\kappa}(x,t)=\exp\left(\frac{\lambda}{2\nu}h_{\kappa}(x,t) \right),
\label{ColeHopf}
\end{equation}
to (\ref{rKPZ}). By It{\^o}'s formula (\ref{rKPZ}) becomes 
\begin{equation}
\frac{\partial Z_{\kappa}(x,t)}{\partial t}=
\nu\frac{\partial^2 Z_{\kappa}(x,t)}
{\partial x^2}+\frac{\lambda\sqrt{D}}{2\nu}\eta_{\kappa}(x,t)Z_{\kappa}
(x,t),
\label{ZKPZ}
\end{equation}
which is a well-defined stochastic differential equation of It{\^o}-type.
Clearly the solution $Z_{\kappa}(x,t)$ to this equation converges to 
that $Z_{\nu,\lambda,D}(x,t)$ of (\ref{SHE}) as $\kappa\to\infty$ 
~\cite{BG1997}. Hence the Cole-Hopf solution (\ref{ColeHopf0}) 
with (\ref{SHE}) can be interpreted as a limit of the solution to the 
equation (\ref{rKPZ}):
\begin{equation}
Z_{\nu,\lambda,D}(x,t)=\lim_{\k\rightarrow\infty}
Z_{\kappa}(x,t).
\end{equation}
Note that if the constant velocity term is absent in~\eqref{rKPZ}, 
there appears an additional term in~\eqref{ZKPZ}, or the equation
(\ref{ZKPZ}) should be understood as representing a stochastic
differential equation of Stratonovich type, which is not well-defined 
in the limit $\kappa\rightarrow\infty$.

The solution of~\eqref{ZKPZ} can be represented as 
the Feynman-Kac formula~\cite{BC1995},
\begin{align}
Z_{\kappa}(x,t)=
\E_x\left(\exp\left[\frac{\lambda\sqrt{D}}{2\nu}
\int_{0}^{t}\eta_{\kappa}\left(b(2\nu s),t-s\right)ds\right]
Z_{\kappa}(b(t),0)\right)e^{-\frac{1}{2}\left(\frac{\lambda\sqrt{D}}{2\nu}
\right)^2C_{\kappa}(0)t},
\label{FKZ}
\end{align}
where $\E_x$ represents the averaging over the standard 
Brownian motion $b(s),~0<s<t$ with $b(0)=x$ and
the initial condition is
\begin{align}
Z_{\kappa}(x,t=0)=
\begin{cases}
0, & x<0,\\
e^{\a B(x)}, & x\ge 0.
\end{cases}
\label{BMin}
\end{align}
Here $B(x)$ represents the one-dimensional Brownian motion
with $B(0)=0$. % and $\\a=(2\nu)^{-3/2}\lambda D^{1/2}$.
Some readers may find 
the following path-integral expression more intuitive.
\begin{equation}
Z_{\kappa}(x,t)=\int_{0}^{\infty}dy
\int_{x(0)=y}^{x(t)=x} D[x(\t)] \exp\left(-S[x(\tau)]+\a B(y)-\frac12
\left(\frac{\lambda\sqrt{D}}{2\nu}
\right)^2 C_{\kappa}(0)t\right),
\label{path}
\end{equation}
where the action $S[x(\t)]$ is given by
\begin{equation}
S[x(\t)]=
\int_0^t d\t\left(
\frac{1}{4\nu}
\left(\frac{dx(\t)}{d\t}\right)^2
-\frac{\lambda\sqrt{D}}{2\nu}\eta_\kappa(x(\t),\t)\right).
\label{action}
\end{equation}

We find that $Z_{\nu,\lambda,D}(x,t)$ defined in~\eqref{ColeHopf0} is 
written in terms of $Z_{\frac12,1,1}(x,t)$  with specific parameters 
$\nu=1/2,\lambda=D=1$ as follows. First using the Feynman-Kac 
formula~\eqref{FKZ}, one has
\begin{align}
&~Z_{\nu,\lambda,D}\left(x,\frac{t}{2\nu}\right) \notag\\
&=\!\lim_{\kappa\rightarrow\infty}
\E_x\!\left(\exp\left[\frac{\lambda\sqrt{D}}{2\nu}
\int_{0}^{\frac{t}{2\nu}}\!\eta_{\kappa}\!\left(\!b(2\nu s),\frac{t}{2\nu}
-s\!\right)ds
-\frac12\left(\frac{\lambda\sqrt{D}}{2\nu}
\right)^2\!C_{\kappa}(0)\frac{t}{2\nu}
\right]\!
Z^{(0)}(b(t),\a)\right),
\label{scaling1}
\end{align}
where we rewrite $Z_{\kappa}(x,0)$ as $Z^{(0)}(x,\a)$
in order to represent explicitly the dependence on $\a$ 
and the independence of $\kappa$.
From the properties of $C_{\kappa}$ written below~\eqref{srandom} 
we find the scaling relations of $C_{\kappa}$ and $\eta_\kappa(x,t)$,
\begin{align}
&C_{\kappa}(x)=aC_{\kappa/a}(ax),\\
&\eta_{\kappa}(x,t)=(ab)^{1/2}\eta_{\kappa/a}(ax,bt),
\label{etas}
\end{align}
where $a,b\in\R$ and the equality ~\eqref{etas} holds in the sense of 
distribution. By use of (\ref{etas}) with $a=1,b=1/2\nu$, 
(\ref{scaling1}) becomes 
\begin{equation}
\lim_{\kappa\rightarrow\infty}\E_x\left(\exp\left[\a
\int_{0}^{t}\eta_{\kappa}\left(b(s),t-s\right)ds-\frac{\a^2}{2}
C_{\kappa}(0)t
\right]
Z^{(0)}(b(t),\a)\right).
\label{scaling2}
\end{equation}
Next one remembers the scaling property of the Brownian motion, 
\begin{align}
a b_x(s)=b_{ax}(a^2s),
\label{sb}
\end{align}
where the equality holds again in the sense of distribution and 
we put the initial position $x$ of $b(s)$ explicitly. 
By (\ref{etas}) and (\ref{sb}) with $a=\a^2, b=\a^4$, 
Eq.~\eqref{scaling2} can be rewritten as
\begin{align}
&\lim_{\kappa\rightarrow\infty}\E_{\a^2 x}\left(\exp\left[
\int_{0}^{\a^4t}\eta_{\kappa/\a^2}\left(b(s),\a^4t-s\right)ds
-\frac{\a^4}{2}
C_{\kappa/\a^2}(0)t
\right]
Z^{(0)}(\a^{-2}b(\a^4 t),\a)\right).
\label{scaling3}
\end{align}
At last noticing from~\eqref{BMin} and~\eqref{sb} that
\begin{align}
Z^{(0)}(\a^{-2}x,\a)=Z^{(0)}(x,\a=1),
\end{align}
we find~\eqref{scaling3} is nothing but 
$Z_{\frac12,1,1}(\a^2x,\a^4t)$.

Thus we have established
\begin{align}
\frac{\lambda}{2\nu}h_{\nu,\lambda,D}\left(x,\frac{t}{2\nu}\right)
=h_{\frac12,1,1}(\a^2x,\a^4t).
\label{generalspecific}
\end{align}
In what follows we restrict our discussions to
$h_{\frac12,1,1}(x,t)$ and $Z_{\frac12,1,1}(x,t)$
(hereafter we omit the indices $\frac12,1,1$).
We remark that, for this special parameter values, 
(\ref{generalgammaX}) reads 
\begin{align}
\gamma_t=\left(\frac{t}{2}\right)^{\frac13},~~
x=2\gamma_t^2X.
\label{specialgammaX}
\end{align}

\subsection{The $\delta$-function Bose gas}
Next we consider the replica partition function 
$\langle Z^N(x,t)\rangle$~($N=0,1,2,\cdots$).
First we perform the Gaussian average over $\eta_{\k}(x,t)$
using the path integral representation~\eqref{path} and~\eqref{srandom}, 
to find 
\begin{align}
\langle e^{N{\int_0}^t d\t\eta(x,\t)}\rangle =e^{\frac12\sum_{j,k=1}^N
C_{\kappa}(x_j-x_k)}.
\end{align}
The right hand side of this equation includes the self-interaction term
$\sum_{j=k=1}^NC_{\kappa}(x_j-x_k)=NC_{\kappa}(0)$, 
which becomes divergent in the limit $\kappa\rightarrow\infty$. 
This term, however, cancels out the last term in~\eqref{path}, 
and thus we can take the limit
$\kappa\rightarrow\infty$.
We obtain
\begin{align}
&~\langle Z^N(x,t)\rangle=\lim_{\k\rightarrow\infty}
\langle Z_{\kappa}^N(x,t)\rangle
\notag\\
&=\prod_{j=1}^N 
\int_{0}^{\infty}dy_j
\int_{x_j(0)=y_j}^{x_j(t)=x}D[x_j(\t)] 
\exp\left[-\int_0^td\t\left(\sum_{j=1}^N\frac12\left(\frac{dx_j(\t)}{d\t}\right)^2
-\sum_{j\neq k=1}^N\delta\left(x_j(\t)-x_k(\t)\right)\right)\right]\notag\\
&~~\hspace{3.5cm}\times\left\langle\exp\left(\sum_{k=1}^NB(y_k)\right)
\right\rangle,
\label{KPZBose}
\end{align}
where $\langle\cdots\rangle$ in the last factor
indicates the remaining average over the Brownian motion $B(y)$.

The right hand side of this equation represents the imaginary-time
dynamics of the $\delta$-function Bose gas with attractive 
interaction with the Hamiltonian $H_N$,
\begin{equation}
H_N=-\frac{1}{2}\sum_{j=1}^N\frac{\partial^2}{\partial x_j^2}
-\frac12\sum_{j\neq k}^N\delta(x_j-x_k),
\label{dBose}
\end{equation}
in terms of which the replica partition function 
$\langle Z^N(x,t)\rangle$ can be written as
\begin{equation}
\langle Z^N(x,t) \rangle=\langle x|e^{-H_Nt}|\Phi\rangle.
\label{KPZdBose1}
\end{equation}
Here $\langle x|$ represents the state with all $N$ particles 
being at the position $x$ and the $|\Phi\rangle$ the initial 
state of the $\delta$-function Bose gas. One can perform 
the average over the Brownian motion $B(y)$ and the dependence 
of $|\Phi\rangle$ on $x_1,\ldots,x_N$ can be explicitly calculated as
\begin{align}
&~~\langle x_1,\cdots,x_N|\Phi\rangle\notag\\
&=\frac{1}{N!}\sum_{P\in S_N}
\left\langle{\exp\left(\sum_{k=1}^N
B(x_{P(k)})\right)}\right\rangle\notag\\
&=\sum_{P\in S_N}
\prod_{j=1}^N\left(\int_{-\infty}^{\infty}
dz_j\frac{e^{-z_j-\frac{(z_j-z_{j-1})^2}{2\left(x_{P(j)}-x_{P(j-1)}\right)}}}
{\sqrt{2\pi (x_{P(j)}-x_{P(j-1)})}}
\Theta\left(x_{P(j)}-x_{P(j-1)}\right)\right)
\notag\\
&=\sum_{P\in S_N}
\prod_{j=1}^Ne^{\frac12(2N-2j+1)x_{P(j)}}\Theta\left(x_{P(j)}-x_{P(j-1)}\right).\label{halfBMic}
\end{align}
Here $S_N$ denotes the set of permutations of order $N$,
$\Theta(x)$ is the step function and we set $x_{P(0)}=0$.
Since we are considering a Boson system, the above function is taken
to be symmetric in $x_1,\cdots,x_N$.

The eigenvalues and eigenfunctions of the $\delta$-Bose gas
can be constructed by using the Bethe 
ansatz~\cite{D2010,Dotsenko2010,CLDR2010,LL1963,M1964}. 
Let $|\Psi_z\rangle$ and $E_z$ be the eigenstate and its eigenvalue
of $H_N$, 
\begin{equation}
H_N|\Psi_z\rangle=E_z|\Psi_z\rangle.
\label{esev}
\end{equation}
By the Bethe ansatz, they are given as
\begin{align}
\langle x_1,\cdots, x_N|\Psi_z\rangle
=C_z\sum_{P\in S_N}{\text{sgn}P}\prod_{1\le j<k\le N}
\left(z_{P(j)}-z_{P(k)}+i\text{sgn}(x_j-x_k)\right) \exp\left(i\sum_{l=1}^N
z_{P(l)}x_l\right)
\label{eigenfunction}
\end{align}
where $C_z$ is the normalization constant, for which a formula 
is given in (\ref{norma}) below. 

For the $\delta$-Bose gas with attractive interaction,
the quasimomenta $z_j~(1\le j\le N)$ which label the state,
are in general complex numbers. $z_j~(1\le j\le N)$ are divided 
into $M$ groups where $1\leq M \leq N$. 
The $\a$th group consists of $n_\a$ quasimomenta
$z_j's$ which share the common real part $q_{\a}$.
Note that 
$\sum_{\a=1}^Mn_\a=N$. 
The quasimomenta in each group line up with
regular intervals with unit length 
along the imaginary direction. 
Using $q_{\a}$ and $n_{\a}~(1\le \a\le M)$, we represent 
$z_j$ $(1\le j\le N)$ as
\begin{align}
&z_j=q_{\a}-\frac{i}{2}\left(n_{\a}+1-2r_\a\right),~~\text{for}~
j=\sum_{\b=1}^{\a-1}n_\b+r_\a,
\label{zq}
\end{align}
where $1\le \a\le M$ and $1\le r_\a \le n_\a$.
The normalization constant $C_z$, which is taken to be a positive real 
number, and the eigenvalue $E_z$ are given by~\cite{Dotsenko2010}
\begin{align}
C_z
&=\left( \frac{\prod_{\a=1}^Mn_\a }{ N!}\prod_{1\le j<k\le N}
\frac{1}{|z_j-z_k-i|^2} \right)^{1/2},
\label{norma}\\
E_z&=\frac12\sum_{j=1}^N z_j^2=\frac{1}{2}\sum_{\a=1}^Mn_\a q_\a^2
-\frac{1}{24}\sum_{\a=1}^M \left(n_\a^3-n_\a\right).
\label{eigenvalue}
\end{align}
There is a problem of completeness of these states. This has not been 
resolved rigorously yet but the fact that one can recover the height 
distribution of the KPZ equation for the narrow wedge initial condition 
is a strong affirmative evidence. Here we proceed assuming its validity 
and will see the consistency with the computations based on the ASEP.
This provides a further evidence that the above Bethe states are in fact 
complete. 

\section{Generating function}
In this section, we give a derivation of the formula~\eqref{KPZfred1}
in Theorem 1. The replica partition function 
$\langle Z^N(x,t)\rangle$~\eqref{KPZdBose1} can be written as
\begin{align}
\langle Z^N(x,t) \rangle&=\int_{-\infty}^{\infty}dy_1\cdots
\int_{-\infty}^{\infty}dy_N
\langle x|e^{-H_Nt}|y_1,\cdots,y_N\rangle\langle y_1,\cdots,y_N|\Phi\rangle.
\label{KPZdBose}
\end{align}
Expanding the propagator $\langle x|e^{-H_Nt}|y_1,\cdots,y_N\rangle$
by the Bethe eigenstates of the $\delta$-Bose gas
(\ref{eigenfunction}), we have
\begin{align}
\langle Z^N(x,t)\rangle
&=
\sum_{M=1}^N\frac{N!}{M!}
\prod_{j=1}^N\int_{-\infty}^{\infty}dy_{j}
\left(\int_{-\infty}^\infty 
\prod_{\a=1}^M\frac{dq_\a}{2\pi}\sum_{n_\a=1}^{\infty} \right)
\delta_{\sum_{\beta=1}^M n_\beta,N} \notag\\
&\quad \times
e^{-E_zt}\langle x|\Psi_z\rangle\langle\Psi_z|y_1,\cdots,y_N\rangle
\langle y_1,\cdots,y_N|\Phi\rangle .
\label{KPZdBose2}
\end{align}
Here we want to perform the integrations over $y_j,~(1\le j\le N)$,
\begin{equation}
\prod_{j=1}^N\int_{-\infty}^{\infty}dy_j\langle\Psi_z|y_1,\cdots,y_N\rangle
\langle y_1,\cdots,y_N|\Phi\rangle
\end{equation}
using (\ref{halfBMic}) and (\ref{eigenfunction}).
But this is not allowed for the moment because 
the integrations over $q_\a,~(1\le \a\le M)$ must be performed 
before those over $y_j,~(1\le j\le N)$. 
To see this explicitly, one uses~\eqref{halfBMic} and 
notices the symmetry of the eigenfunction, 
$\langle\Psi_z|y_{P(1)},\cdots,y_{P(N)}\rangle
=\langle\Psi_z|y_{1},\cdots,y_{N}\rangle$, to find that 
the right hand side of~\eqref{KPZdBose2} is represented as
\begin{align}
\langle {Z}^N(x,t)\rangle
&=\sum_{M=1}^N\frac{N!}{M!}\prod_{j=1}^N\int_{y_{j-1}}^{\infty}dy_{j}
\prod_{\a=1}^M\left(
\int_{-\infty}^\infty\frac{dq_\a}{2\pi}
\sum_{n_\a=1}^{\infty}\right)\delta_{\sum_{\beta=1}^Mn_\beta,N}
\notag\\
&~\times e^{-E_zt}
\langle x|\Psi_z\rangle\langle\Psi_z|y_1,\cdots,y_N\rangle
\prod_{j=1}^Ne^{\frac12(2N-2j+1)y_j}.
\label{Betheexpand}
\end{align}
It is clear that the integrand on the right hand side of this equation
is not integrable on $y_j~(1\le j\le M)$ due to the factor
$\exp((2N-2j+1)y_j)$ while it is integrable on $q_\a$~$(1\le \a\le M)$
thanks to the factor $e^{-n_\a q_\a^2t/2}$ in $e^{-E_zt}$.

\subsection{Deformation of contours}
To exchange the order of integrations, we notice 

\vspace{3mm}
\noindent
{\bf Lemma 4}
{\it 
In~\eqref{Betheexpand}, we can deform the contour of
$q_\a~(1\le \a\le M)$ to $\R-ic$ where $c$ is an arbitrary real 
constant.
}

\vspace{3mm}
\noindent
{\bf Proof}
In~\eqref{Betheexpand}, we change the variables 
from $q_\a$ to $u_\a~(1\le j\le M)$ such that
\begin{equation}
u_\a=q_\a-q_{\a+1},~(1\le \a \le M-1),~~u_M=q_M.
\end{equation}
The singularity in $u_{\a}$ of the integrand
in~\eqref{Betheexpand} comes only from the factor $|C_z|^2$.  
Note that it depends only on $u_1,\cdots, u_{M-1}$
and is independent of $u_M$. Thus there are no poles of
$u_M$ in~\eqref{Betheexpand}, and we can deform the 
contour of $u_M$ to $\R-ic$. Noting that
\begin{equation}
\prod_{\a=1}^{M-1}
\int_{-\infty}^\infty\frac{du_\a}{2\pi}
\times \int_{\R-ic} \frac{du_M}{2\pi}
=\prod_{\a=1}^M
\int_{\R-ic}\frac{dq_\a}{2\pi},
\end{equation}
we find the statement holds.
\begin{flushright}
$\blacksquare$
\end{flushright}

\vspace{3mm}

\noindent
Using Lemma 4, we deform the contours of $q_\a,~(1\le \a\le M)$
to $\R - ic$ where $c$ is a positive constant so large that 
the factor
$\exp\left(-i\sum_{j=1}^Nz^*_{P(j)}y_j+\frac12\sum_{j=1}^N
(2N-2j+1)y_j\right)$ 
in the integrand of~\eqref{Betheexpand}
converge when $y_j>0$ goes to infinity.
Here $z_j^*=q_{\a (j)}+\frac{i}{2}\left(n_{\a(j)}+1-2r(j)\right)$
but one has to be careful about the notation. 
This is a usual complex conjugate of $z_j$ before lemma 4, i.e., when
$q_{\a(j)}$ is real. But now it is not
because after the deformation of the contour, $q_{\a(j)}$ has 
an imaginary part.

After this deformation, we can perform the integrations of 
$y_j~(1\le j \le N)$ before those of $q_\a~(1\le \a \le M)$.
Eq.~\eqref{KPZdBose2} can now be expressed as
\begin{align}
\langle {Z}^N(x,t)\rangle
=\sum_{M=1}^N\frac{1}{M!}
\prod_{\a=1}^M\left(
\int_{\R-ic}\frac{dq_\a}{2\pi}
\sum_{n_\a=1}^{\infty}\right)\delta_{\sum_{\beta=1}^Mn_\beta,N}
\langle x|\Psi_z\rangle\langle\Psi_z|\Phi\rangle
e^{-E_zt}.
\label{afterdeform}
\end{align}
Here $\langle x|\Psi_z\rangle$ is given by~\eqref{eigenfunction} 
with $x_1=\cdots=x_N=x$ and
$\langle\Psi_z|\Phi\rangle$ is computed as 
\begin{align}
&~\langle \Psi_z|\Phi\rangle
=\prod_{j=1}^N\int_{-\infty}^{\infty}dy_j\langle\Psi_z|y_1,\cdots,y_N\rangle
\langle y_1,\cdots,y_N|\Phi\rangle
\notag\\
&=C_z
\sum_{P\in S_N}\text{sgn}P
\prod_{l=1}^{N}\int_{y_{l-1}}^{\infty}dy_{l}
e^{-i\left(z^*_{P(l)}+\frac{i}{2}(2N-2l+1)\right)y_{l}}\prod_{1\le j<k\le N}
\left(z^*_{P(j)}-z^*_{P(k)}+i)\right)\notag\\
&=N!C_z \sum_{P\in S_N}{\text{sgn}P}\prod_{1\le j<k\le N}
\left(z^*_{P(j)}-z^*_{P(k)}+i\right) 
\prod_{l=1}^N\frac{-1}{-i(z^*_{P(N)}+\cdots +z^*_{P(N-l+1)})
+l^2/2}.
\label{nodangerous}
\end{align}
In (\ref{afterdeform}), we take the imaginary part $c$ of $q_{\a}$ in
such a way 
that we can perform the integrations of $y_j~(1\le j\le N)$ 
in~\eqref{nodangerous}, i.e. $\text{Re}(-i(z^*_{P(N)}+\cdots +z^*_{P(N-l+1)})
+l^2/2)<0$ for any $l~(1\le l\le N)$. For example, if we fix $c$ such that
$c>N/2+\max_{\a}n_\a/2$, the above condition is satisfied.

\subsection{Combinatorial identities}
For further analysis of the integrand in~\eqref{afterdeform},
we need two combinatorial identities for $\langle x|\Psi_z\rangle$ and
$\langle\Psi_z|\Phi\rangle$. The first one is for
$\langle x|\Psi_z\rangle$. One has 
\begin{equation}
\sum_{P\in S_N}{\text{sgn}P}\prod_{1\le j<k \le N}
\left(w_{P(j)}-w_{P(k)}+if(j,k)\right)
=N!\prod_{1\le j<k\le N}(w_j-w_k)
\label{ci1}
\end{equation}
for any complex variables $w_j$~$(1\le j \le N)$ and 
$f(j,k)$. This identity was derived as Lemma~1 in \cite{PS2010p}.

The next one for the term $\langle\Psi_z|\Phi\rangle$ is 

\vspace{3mm}
\noindent
{\bf Lemma 5}
{\it 
For any complex numbers $w_j~(1\le j\le N)$ and $a$,  
\begin{align}
&~\sum_{P\in S_N}{\text{\rm sgn} P}\prod_{1\le j<k \le N}
\left(w_{P(j)}-w_{P(k)}+a\right) 
\prod_{m=1}^N\frac{1}{w_{P(N)}+\cdots +w_{P(N-m+1)}
+m^2a/2}
\notag\\
&~=\prod_{1\le j<k\le N}(w_j-w_k)\prod_{m=1}^N\frac{1}{w_m+a/2}.
\label{ci2}
\end{align}
}

\noindent
{\bf Proof}
A similar identity appears in the context of ASEP with the step
Bernoulli initial condition and has been proved in section III 
of~\cite{TW2009}.
Here we follow the same strategy using mathematical induction. 
Let us call the left hand side and the right hand side of (\ref{ci2}) as 
$\psi_N$ and $\phi_N$ respectively, i.e., 
\begin{align}
&~\psi_N(w_1,\cdots,w_N):=
\sum_{P\in S_N}{\text{sgn} P}\prod_{1\le j<k\le N}(w_{P(j)}-w_{P(k)}+a)\notag\\
&\hspace{5.5cm}\times\prod_{m=1}^N
\frac{1}{w_{P(N)}+\cdots+w_{P(N-m+1)}+m^2a/2},
\label{phin}
\\
&~\phi_N(w_1,\cdots,w_N):=
\prod_{1\le j<k\le N}(w_j-w_k)\prod_{m=1}^N\frac{1}{w_m+a/2}.
\label{phin2}
\end{align}
We want to show $\psi_N=\phi_N$. 
We easily see that it holds for the case $N=1$. 
Let us assume that it holds for $N-1$ case.
In~\eqref{phin}, we first sum over all permutations
with $P(1)=l$ fixed and then sum over $l$.
Noticing $\prod_{1 < k\le N}(w_l-w_{P(k)}+a)
=\prod_{k(\neq l)}(w_l-w_k+a)$,
we see
\begin{align}
&~\psi_N(w_1,\cdots,w_N)\notag\\
&=
\frac{1}{\sum_{j=1}^Nw_j+N^2a/2}
\sum_{l=1}^N(-1)^{l+1}\prod_{j(\neq l)}(w_l-w_j+a)
\psi_{N-1}(w_1,\cdots,w_{l-1},w_{l+1},\cdots,w_N)\notag\\
&=
\frac{1}{\sum_{j=1}^Nw_j+N^2a/2}
\sum_{l=1}^N(-1)^{l+1}\prod_{j(\neq l)}(w_l-w_{j}+a)
\phi_{N-1}(w_1,\cdots,w_{l-1},w_{l+1},\cdots,w_N),
\end{align}
where we used the assumption of the mathematical induction,
$\psi_{N-1}=\phi_{N-1}$, in the second equality.
The identity $\psi_N=\phi_N$ is now equivalent to
\begin{align}
\sum_{l=1}^{N}(-1)^{l+1}\prod_{j(\neq l)}(w_l-w_j+a)
\frac{\phi_{N-1}(w_1,\cdots,w_{l-1},w_{l+1},\cdots,w_N)}
{\phi_N(w_1,\cdots,w_N)}=\sum_{j=1}^Nw_j+\frac{N^2a}{2}.
\label{iden1}
\end{align}
From the definition~\eqref{phin2} of $\phi_N$, we get
\begin{align}
\phi_{N-1}(w_1,\cdots,w_{l-1},w_{l+1},\cdots,w_N)
&=
\frac{w_l+a/2}{\prod_{j=1}^{l-1}(w_j-w_l)\cdot\prod_{j=l+1}^{N}(w_l-w_j)}
\phi_N(w_1,\cdots,w_N)\notag\\
&=
\frac{w_l+a/2}
{(-1)^{l-1}\prod_{j(\neq l)}(w_l-w_j)}
\phi_N(w_1,\cdots,w_N).
\end{align}
Substituting this equation into~\eqref{iden1},
one sees that it is now enough to show
\begin{equation}
\sum_{l=1}^N(w_l+a/2)\prod_{j(\neq l)}\left(
1+\frac{a}{w_l-w_j}\right)
=\sum_{j=1}^Nw_j+\frac{N^2a}{2}.
\label{etoshow}
\end{equation}
This is proved as follows. One notices that 
the left hand side can be represented by a contour integral,
\begin{equation}
\frac{1}{2\pi ia}\int_{C_R}dz(z+a/2)
\prod_{j=1}^N\left(1+\frac{a}{z-w_j}\right),
\label{ci}
\end{equation}
where $C_R$ is a contour enclosing the origin anticlockwise
with radius $R$ taken to be so large that the contour surrounds
all the poles $w_j~(1\le j\le N)$ in the integrand. 
The contour integration can be performed as follows.
Expanding the product in the integrand, one finds
\begin{gather}
\frac{1}{2\pi ia}\int_{C_R}dz
\left[
\left( z+\frac{a}{2} \right)
+\sum_{l=1}^N\frac{a}{z-w_l}
\left( z+\frac{a}{2} \right)
+\sum_{1\le l<m\le N}\frac{a^2}{(z-w_l)(z-w_m)}
\left( z+\frac{a}{2} \right) + \cdots  \right. \notag\\
\quad \left.
+ ~a^N ~\prod_{j=1}^N \frac{1}{z-w_j}\left(z+\frac{a}{2}\right)
\right].
\label{expand}
\end{gather}
Further expanding the second and third terms in the integrand in $1/z$, we see
\begin{align}
&\frac{a}{z-w_l}
\left(
z+\frac{a}{2}
\right)=a+\frac{a(w_l+a/2)}{z}+O\left(\frac{1}{z^2}\right),\notag\\
&\frac{a^2}{(z-w_l)(z-w_m)}
\left(
z+\frac{a}{2}
\right)=\frac{a^2}{z}+O\left(\frac{1}{z^2}\right).
\end{align}
The higher terms in \eqref{expand} are of order $O(1/z^2)$ 
and thus do not contribute to the contour integral.
Hence \eqref{ci} is calculated as
\begin{align}
&~\frac{1}{2\pi ia}\int_{C_R}dz (z+a/2)
\prod_{j=1}^N\left(1+\frac{a}{z-w_j}\right)\notag\\
&=\frac{1}{2\pi ia}\int_{C_R}dz 
\left(\sum_{l=1}^N\frac{a(w_l+a/2)}{z}+\sum_{1\le l<m\le N}\frac{a^2}{z}\right)\notag\\
&=\sum_{l=1}^N\left(w_l+\frac{a}{2}\right)+\sum_{l<m}a=\sum_{j=1}^Nw_j
+\frac{N^2a}{2},
\end{align}
thus we obtain~\eqref{etoshow}.
\begin{flushright}
$\blacksquare$
\end{flushright}

Using the identities~\eqref{ci1} and~\eqref{ci2} to~\eqref{eigenfunction} 
and~\eqref{nodangerous} respectively, 
we get
\begin{align}
&\langle x|\Psi_z\rangle
=N!C_z\prod_{1\le j<k\le N}(z_j-z_k)e^{i\sum_{l=1}^Nz_lx},
\label{Psicompact}
\\
&\langle\Psi_z|\Phi\rangle
=
i^{-N}N!C_z\prod_{1\le j<k\le N}(z^*_j-z^*_k)\prod_{l=1}^N\frac{1}{z^*_l+i/2}.
\end{align} 
Thus using these relations and~\eqref{norma}, we find that the factor
$\langle x|\Psi_z\rangle\langle\Psi_z|\Phi\rangle$
in~\eqref{afterdeform} becomes
\begin{align}
\langle x|\Psi_z\rangle\langle\Psi_z|\Phi\rangle
=N!\prod_{\a=1}^M\frac{(n_{\a}!)^2}{n_{\a}}
\prod_{1\le j<k\le N}\frac{|z_j-z_k|^2}{|z_j-z_k-i|^2}
\prod_{l=1}^N\frac{e^{iz_lx}}{iz^*_l-1/2},
\label{integrand0}
\end{align}
Here we used the fact that $z_j^*-z_k^*$ is the complex conjugate
of $z_j-z_k$ although $z_j^*$ is not that of $z_j$ as mentioned below
Lemma 4. The common imaginary part $c$ of $q_{\a(j)}$ and $q_{\a(k)}$
cancels out by the subtraction. 

We want to rewrite this equation in terms of $q_{\a}$ and $n_{\a}$ 
in~\eqref{zq}.
For the last factor of this equation, we easily find 
\begin{align}
\prod_{l=1}^N\frac{e^{iz_lx}}{iz^*_l-1/2}=\prod_{\a=1}^Me^{in_\a q_\a x}
\prod_{r=1}^{n_{\a}}\frac{1}{iq_{\a}+\frac12(n_{\a}-2r)}.
\end{align} 
From~\eqref{norma} and~\eqref{Psicompact}, we know the remaining factors 
in~\eqref{integrand0} is represented by $|\langle 0|\Psi_z\rangle|^2$.
For this quantity, the following result was obtained in Appendix B 
in~\cite{Dotsenko2010},
\begin{align}
|\langle 0|\Psi_z\rangle|^2&=N!\prod_{\a=1}^M\frac{(n_{\a}!)^2}{n_{\a}}
\prod_{1\le j<k\le N}\frac{|z_j-z_k|^2}{|z_j-z_k-i|^2}\notag\\
&=\frac{N!}{\prod_{\a=1}^Mn_{\a}}
\prod_{1\le\a<\b\le M}\frac{|q_\a-q_\b-\frac{i}{2}(n_{\a}-n_{\b})|^2}
{|q_\a-q_\b-\frac{i}{2}(n_{\a}+n_{\b})|^2}.
\label{Psi0}
\end{align}
From~\eqref{integrand0}--\eqref{Psi0}, we get
\begin{align}
&~\langle x|\Psi_z\rangle\langle\Psi_z|\Phi\rangle\notag\\
&=
N!
\prod_{\a<\b}^M
\frac{|q_\a-q_\b-\frac{i}{2}(n_\a-n_\b)|^2}
     {|q_\a-q_\b-\frac{i}{2}(n_\a+n_\b)|^2}
\prod_{\a=1}^M\frac{e^{in_{\a}q_{\a}x}}{n_\a}\prod_{r=1}^{n_\a}
\frac{1}{iq_\a+\frac{1}{2}(n_\a-2r)}.
\label{integrand1}
\end{align}

We can further deform~\eqref{integrand1} to an
expression in terms of a determinant by 
using the Cauchy's determinant formula,
\begin{equation}
\frac{\prod_{\a<\b}^M(a_\a-a_\b)(b_\a-b_\b)}
{\prod_{\a,\b=1}^M(a_{\a}-b_\b)}=(-1)^{\frac{M(M-1)}{2}}
\det\left(\frac{1}{a_\a-b_\b}\right),
\end{equation}
and a few basic properties of determinant. We find
\begin{align}
&~~\langle x|\Psi_z\rangle\langle\Psi_z|\Phi\rangle\notag\\
&=2^MN!\det\left(\frac{1}{n_j+n_k+2i(q_j-q_k)}\right)_{j,k=1}^M
\prod_{\a=1}^Me^{in_{\a}q_\a x}\prod_{r=1}^{n_\a}
\frac{1}{iq_\a+\frac{1}{2}(n_\a-2r)}
\notag\\
&=2^MN!\prod_{\a=1}^M\left(\int_0^{\infty}d\o_\a\right)
\det\left(e^{-\o_j(n_j+n_k+2i(q_j-q_k))}\right)_{j,k=1}^M
\prod_{\a=1}^Me^{in_{\a}q_\a x}\prod_{r=1}^{n_\a}
\frac{1}{iq_\a+\frac{1}{2}(n_\a-2r)}
\notag\\
&=2^MN!\prod_{\a=1}^M \left(\int_0^{\infty}d\o_\a\right)
\det\left(e^{in_jq_jx-n_j(\o_j+\o_k)-2iq_j(\o_j-\o_k)}
\prod_{r=1}^{n_j}
\frac{1}{iq_j+\frac{1}{2}(n_j-2r)}
\right)_{j,k=1}^M,
\label{integrand2}
\end{align}
where in the last equality, we used a simple fact
\begin{equation}
 \det(a_j^{b_j+b_k}) = \det((a_j a_k)^{b_j}).
\end{equation}

\subsection{Fredholm determinant representation of 
the generating function}
From~\eqref{eigenvalue},~\eqref{afterdeform}, 
and~\eqref{integrand2},
we obtain an expression of $\langle Z^N(x,t)\rangle$ in terms of
the determinant,
\begin{align}
&\quad\langle Z^N(x,t)\rangle e^{\frac{Nt}{24}+\frac{Nx^2}{2t}}
\notag\\
&=\sum_{M=1}^N\frac{2^MN!}{M!}\prod_{\a=1}^M
\left(\sum_{n_\a=1}^{\infty}
\int_{\R-ic}\frac{dq_\a}{2\pi}
e^{-\frac{t}{2}n_\a q_\a^2+\frac{t}{24}n_\a^3}
\int_0^\infty d\o_\a\right)
\delta_{\sum_{\beta=1}^Mn_\beta,N}\notag\\
&\hspace{7.5cm}\times
\det\left(\frac{e^{in_jq_jx+n_j\frac{x^2}{2t}-n_j(\o_j+\o_k)-2iq_j(\o_j-\o_k)}}
{\prod_{r=1}^{n_j}{iq_j+\frac{1}{2}(n_j-2r)}}
\right)_{j,k=1}^M\notag\\
&=\sum_{M=1}^N\frac{N!}{M!}\prod_{\a=1}^M\left(
\int_0^\infty d\o_\a\sum_{n_\a=1}^{\infty}\right)\delta_{\sum_{\beta=1}^M
n_\beta,N}\notag\\
&\hspace{4cm}\times\det\left(\int_{\R-ic}\frac{dq}{\pi}
\frac{e^{in_jq_jx+n_j\frac{x^2}{2t}-\frac{t}{2}n_jq^2+\frac{t}{24}n_j^3-n_j(\o_j+\o_k)-2iq(\o_j-\o_k)}}
{\prod_{r=1}^{n_j}{iq+\frac{1}{2}(n_j-2r)}}
\right)_{j,k=1}^M.
\label{zndet}
\end{align}
The imaginary part $c$ of the contour of $q$ is discussed 
below~\eqref{afterdeform} 
and we find
it is sufficient to satisfy the condition $c>N/2+\max_{\a}n_\a/2$.
Note that in~\eqref{zndet}, the contour satisfying this condition 
passes below the poles
$q=j-n_\a/2~(j=0,1,\cdots,[n_{\a}/2])$ where $[a]$ is the largest integer
which is smaller than $a$ and that we can deform the contour as long as 
it satisfies this property. Thus in \eqref{zndet}, we can relax 
the condition to $c>\max_{\a}n_\a/2$.

Substituting this equation into~\eqref{gr}, we eventually
get the Fredholm determinant representation of the generating 
function~\eqref{KPZfred1}.
Remembering (\ref{specialgammaX}),
we have
\begin{align}
&~G_{\gamma_t}(s;X) %:=\sum_{N=0}^\infty\frac{(-e^{-\gamma_t s})^N}
\notag\\
&=1+\sum_{N=1}^\infty\sum_{M=1}^N\frac{(-e^{-\gamma_t s})^N}{M!}
\prod_{\a=1}^M\left(
\int_0^\infty d\o_\a
\sum_{n_\a=1}^{\infty}\right)\delta_{\sum_{\beta=1}^Mn_\beta,N}
\notag\\
&\hspace{3.5cm}~\times\det\left(\int_{\R-ic}\frac{dq}{\pi}
\frac{
e^{2i n_j\gamma_t^2 qX+n_j\gamma_tX^2-\gamma_t^3n_jq^2+\frac{\gamma_t^3}{12}n_j^3-n_j(\o_j+\o_k)-2iq(\o_j-\o_k)}}
{{\prod_{r=1}^{n_j}
{(iq+\frac{1}{2}(n_j-2r))}}}
\right)_{j,k=1}^M\notag\\
&=\sum_{M=0}^\infty\frac{(-1)^M}{M!}
\prod_{\a=1}^M
\int_0^\infty d\o_\a\notag\\
&\hspace{1cm}\times\det\left(
\sum_{n=1}^{\infty}(-1)^{n-1}
\int_{\R-ic_n}\frac{dq}{\pi}\frac{
e^{2i nq\gamma_t^2X+n\gamma_tX^2-\gamma_t^3nq^2+\frac{\gamma_t^3}{12}n^3-n(\o_j+\o_k)-2iq(\o_j-\o_k)
-\gamma_t ns}}
{{\prod_{r=1}^{n}
{(iq+\frac{1}{2}(n-2r))}}}
\right)_{j,k=1}^M
\label{KPZfred2},
\end{align}
with $c_n>n/2$.
Shifting the variable $q$ to $q+iX/\gamma_t$, we get
\begin{align}
&~G_{\gamma_t}(s;X)\notag\\
&=\sum_{M=0}^\infty\frac{(-1)^M}{M!}
\prod_{k=1}^M
\int_0^\infty d\o_k\notag\\
&\hspace{1cm}\times\det\left(
\sum_{n=1}^{\infty}(-1)^{n-1}
\int_{\R-ic_n}\frac{dq}{\pi}\frac{
e^{-n(\o_j+\o_k)-2iq(\o_j-\o_k)-\gamma_t^3nq^2+\frac{\gamma_t^3}{12}n^3
-\gamma_t ns}}
{{\prod_{r=1}^{n}
{(iq-\frac{X}{\gamma_t}+\frac{1}{2}(n-2r))}}}
\right)_{j,k=1}^M
\label{KPZshift},
\end{align}
with $c_n>X/\gamma_t+n/2$.
Applying the relation
\begin{equation}
\prod_{r=1}^n\frac{1}{iq-X/\gamma_t+\frac{1}{2}(n-2r)}=
\frac{\Gamma\left(iq-\frac{X}{\gamma_t}-\frac{n}{2}\right)}
{\Gamma\left(iq-\frac{X}{\gamma_t}+\frac{n}{2}\right)}
\end{equation}
to~\eqref{KPZshift},
we finally obtain~\eqref{KPZfred1}.

\section{Another expression of the kernel}
In this section we give a derivation of the 
expression~\eqref{alternative1} in
Proposition 2.
We use the following two relations,

\vspace{3mm}
\noindent
{\bf Lemma 6}

\noindent
(a)
{\it 
We set $a\in\R$ and ~$m,~n\ge0$. 
When ${\rm Im}~q<-n/2+a$, we have
\begin{align}
\frac{\Gamma\left(iq+a-\frac{n}{2}\right)}{\Gamma\left(iq+a+\frac{n}{2}\right)} e^{\frac{m^3n^3}{3}}
=\int_{-\infty}^{\infty}dy
\Ai_{\Gamma}^\Gamma\left(y,\frac{1}{2m},iq+a\right)
e^{mny},
\label{lemma6(1)}
\end{align}
where
\begin{align}
\Ai_{\Gamma}^\Gamma\left(a,b,c\right)=\frac{1}{2\pi}
\int_{\Gamma_{i\frac{c}{b}}}dz e^{iaz+iz^3/3}
\frac{\Gamma(ibz+c)}{\Gamma(-ibz+c)}
\label{AiGG}
\end{align}
with $\Gamma_p$ defined in~\eqref{AiG2} below.

\noindent
{\rm (b)} For $u,v,x\in\R$ and $w\ge 0$, we have
\begin{align}
&~\frac{1}{2\pi}\int_{-\infty}^{\infty}dp
\Ai_{\Gamma}^\Gamma\left(p^2+v,w,iwp+u\right)e^{ipx}\notag\\
&=
\frac{1}{2^{\frac{1}{3}}}
\Ai^\Gamma\left(2^{-\frac{2}{3}}(v+x),2^{\frac{1}{3}}w,u\right)
\Ai_\Gamma\left(2^{-\frac{2}{3}}(v-x),2^{\frac{1}{3}}w,u\right),
\label{lemma6(2)}
\end{align}
where $\Ai^\Gamma(a,b,c)$ and $\Ai_\Gamma(a,b,c)$ are defined by~\eqref{AiG1} 
and~\eqref{AiG2} respectively.
}

\vspace{3mm}
\noindent
{\bf Proof}

\noindent
(a)
The right hand side of~\eqref{lemma6(1)} is written as
\begin{align}
\frac{1}{2\pi}\int_{-\infty}^{\infty}dy\int_{\R+imn}dz
\frac{\Gamma\left(i\frac{z}{2m}+iq+a\right)}
{\Gamma\left(-i\frac{z}{2m}+iq+a\right)}
e^{i(z-imn)y+i\frac{z^3}{3}}.
\end{align}
Here we used the fact that 
the contour of $z$ can be deformed from $\Gamma_{2m(ia-q)}$ to 
$\R+imn$ since the imaginary part of the poles 
$2m(ia-q+ir)$, $r=0,1,2,\cdots$ of the integrand is larger than
$mn$ when the condition $\text{Im~}q<a-n/2$ is satisfied. In this
equation, we change the variable $z$ on the right hand side to $y_2=z-imn$
and get
\begin{align}
&~\frac{1}{2\pi}
\int_{-\infty}^{\infty}dy\int_{-\infty}^{\infty}d{y}_2 
\frac{\Gamma\left(iq+a-\frac{n}{2}+i\frac{{y}_2}{2m}\right)}
{\Gamma\left(iq+a+\frac{n}{2}-i\frac{{y}_2}{2m}\right)}
e^{iy{y}_2+i\frac{({y}_2+im n)^3}{3}}\notag\\
&=\int_{-\infty}^{\infty}d{y}_2\delta({y}_2)
\frac{\Gamma\left(iq+a-\frac{n}{2}+i\frac{{y}_2}{2m}\right)}
{\Gamma\left(iq+a+\frac{n}{2}-i\frac{{y}_2}{2m}\right)}
e^{i\frac{({y}_2+im n)^3}{3}}=\frac{\Gamma\left(iq+a-\frac{n}{2}\right)}
{\Gamma\left(iq+a+\frac{n}{2}\right)}
e^{\frac{m^3 n^3}{3}}.
\end{align}
\noindent
(b)
The left hand side of~\eqref{lemma6(2)} reads
\begin{align}
\frac{1}{(2\pi)^2}\int_{-\infty}^{\infty} dp
\int_{\Gamma_{-p+i\frac{u}{w}}}dz e^{i(p^2+v)z+i\frac{z^3}{3}+ipx}
\frac{\Gamma\left(iw(z+p)+u\right)}
{\Gamma\left(-iw(z-p)+u\right)}.
\end{align}
By applying the change of variables $p=(z_1-z_2)/2^{2/3}$
and $z=(z_1+z_2)/2^{2/3}$, we obtain the desired expression.
\begin{flushright}
$\blacksquare$
\end{flushright}

Using Lemma~6, we can obtain~\eqref{alternative1}.
Applying (a) of the lemma with $a=-X/\gamma_t$ and $m=\gamma_t/2^{2/3}$ 
to~\eqref{th3kernel1}, one has
\begin{align}
K_X(\o_j,\o_k)
&=\sum_{n=1}^{\infty}(-1)^{n-1}
\int_{\R-ic_n}\frac{dq}{\pi}
\int_{-\infty}^{\infty}dy_1
\Ai_{\Gamma}^\Gamma\left(y_1,\frac{1}{2^{\frac{1}{3}}\gamma_t},iq-\frac{X}{\gamma_t}
\right)
e^{-2iq(\o_j-\o_k)}\notag\\
&
\hspace{5cm}\times e^{-n(\o_j+\o_k+\gamma_t^3q^2+\gamma_ts-2^{-\frac{2}{3}}\gamma_ty_1)}.
\end{align}
At this point, the contour of $q$ can be replaced to $\R$ since
since the contour of $z$ in the definition of 
$A^{\Gamma}_\Gamma(a,b,c)$~\eqref{AiGG} pass below the singularity
of the Gamma function. 
Changing the variables $q,~\o_j$ and $y_1$ to 
$p=2^{1/3}\gamma_t q,~\xi_j=2\gamma_t^{-1}\o_j$ and
$y=y_1/2^{2/3}-p^2/2^{2/3}-(\xi_j+\xi_k)/2$,
we see 
$K_X(\o_j,\o_k)d\o_j=\bar{K}_X(\xi_j,\xi_k)d\xi_j$, where
\begin{align}
\bar{K}_X(\xi_j,\xi_k)
&=\frac{2^{\frac{1}{3}}}{2\pi}
\int_{-\infty}^{\infty}dy
\int_{-\infty}^{\infty}{dp}
\Ai_{\Gamma}^\Gamma
\left(p^2+2^{\frac{2}{3}}y+\frac{\xi_j+\xi_k}{2^{\frac{1}{3}}},
\frac{1}{2^{\frac{1}{3}}\gamma_t},\frac{ip-2^{\frac{1}{3}}X}{2^{\frac{1}{3}}
\gamma_t}
\right)
e^{-i\frac{(\xi_j-\xi_k)}{2^{\frac{1}{3}}}p}
\notag\\
&\hspace{5cm}
\times 
\sum_{n=1}^{\infty}(-1)^{n-1}e^{-\gamma_tn(s-y)}.
\end{align}
Thus we can apply (2) of Lemma~6 to this equation 
with $v=2^{2/3}y+(\xi_j+\xi_k)/2^{1/3}$, $w=1/(2^{1/3}\gamma_t)$, 
$u=-X/\gamma_t$, $x=(\xi_j-\xi_k)/2^{1/3}$
and arrive at~\eqref{alternative1}. 

\section{Multi-point distribution}
\subsection{Generating function}
We consider the $n$-point generating function defined 
by~\eqref{multdist} and derive the result \eqref{apge}.
Eq.~\eqref{multdist} can be expanded in terms of the
replica partition function 
$\langle Z(x_{l_1},t)\cdots Z(x_{l_N},t)\rangle$, 
\begin{align}
G_{\gamma_t}(\{s\}_n,\{X\}_n)&=1+\sum_{N=1}^{\infty}\frac{(-1)^Ne^{\frac{tN}{24}}}{N!}
\notag\\
&\hspace{1cm}\times
\sum_{l_1,\cdots,l_N=1}^ne^{-\gamma_t\sum_{j=1}^N(s_j-X_j^2)}
\langle Z(2\gamma_t^2X_{l_1},t)\cdots Z(2\gamma_t^2X_{l_N},t)\rangle.
\end{align}
As we discussed in Sec. 3 and 4, we can express 
the replica partition function 
in terms of the Bethe states~\eqref{eigenfunction}--\eqref{eigenvalue} 
of the $\delta$-Bose gas. Specifically we can the generalize 
\eqref{afterdeform} to the case where $x'$s are distinct and
we get
\begin{align}
&~\langle Z(x_{l_1},t)\cdots Z(x_{l_N},t)\rangle\notag\\
&=\sum_{M=1}^N\frac{1}{M!}
\prod_{\a=1}^M\left(
\int_{\R-ic}\frac{dq_\a}{2\pi}
\sum_{n_\a=1}^{\infty}\right)\delta_{\sum_{\beta=1}^Mn_\beta,N}
\langle x_{l_1},\cdots,x_{l_N}|
\Psi_z\rangle\langle\Psi_z|\Phi\rangle
e^{-E_zt}.
\label{afterdeform2}
\end{align}
Thus the generating function
is written as
\begin{align}
G_{\gamma_t}(\{s\}_n,\{X\}_n)=1+\sum_{N=1}^{\infty}\frac{(-1)^Ne^{\frac{tN}{24}}}{N!}
\prod_{\a=1}^M\left(
\int_{\R-ic}\frac{dq_\a}{2\pi}
\sum_{n_\a=1}^{\infty}\right)\delta_{\sum_{\b=1}^Mn_\b,N}
A_zB_z
\label{AzBz}
\end{align}
where 
\begin{align}
&A_z=e^{-E_zt}\langle 0|\Psi_z\rangle\langle\Psi_z|\Phi\rangle,\\
&B_z=\sum_{l_1,\cdots,l_N=1}^ne^{-\gamma_t\sum_{j=1}^n(s_{l_j}-X_{l_j}^2)}
\frac{\langle 2\gamma_t^2X_{l_1},\cdots,2\gamma_t^2X_{l_N}|\Psi_z\rangle}
{\langle 0|\Psi\rangle}.
\label{psapprox}
\end{align}
Note that in two factors $A_z$ and $B_z$, only $A_z$ depends on the 
initial condition and we can readily use~\eqref{eigenvalue} and~\eqref{integrand2} for $A_z$.
On the other hand, the factor $B_z$, which
is irrelevant to specific initial conditions,
has already appeared in the narrow wedge
initial condition and the following approximation 
has been proposed in~\cite{PS2010p,PS2010p2}.

For $B_z$, we have to estimate the equation with the following type,
\begin{align}
\sum_{l_1,\cdots,l_N=1}^n
\langle x_{l_1},\cdots,x_{l_N}|\Psi_z\rangle\prod_{j=1}^Ne^{g_{l_j}}.
\label{sumeigen}
\end{align}
Here $g_{j}~(j=1,\cdots,n)$ are variables which depend on $j$. 
Although the eigenfunction $\langle x_{l_1},\cdots,x_{l_N}|\Psi_z\rangle$
is given in~\eqref{eigenfunction}, 
the following compact form given in~\cite{Dotsenko2010} is convenient
for our current discussion,
\begin{align}
\langle y_{1},\cdots,y_{N}|\Psi_z\rangle
=
\sum_{P\in S_N}\!\!\!{}^{^{'}}
A_P(y_1,\cdots,y_N)
\exp\left(i\sum_{\a=1}^Mq_{\a}
\sum_{c\in\Omega_\a(P)}y_c-\frac{1}{4}\sum_{\a=1}^M\sum_{c,c'\in\Omega_\a(P)}
|y_c-y_{c'}|
\right),
\label{eigen2}
\end{align}
where
$A_P$,  $\Omega_\a(P)$ and the prime in the summation 
are defined as follows: The amplitude $A_p$ is represented as 
\begin{align}
A_P(y_1,\cdots,y_N)=\text{sgn}P\cdot C_z\prod_{\a=1}^M\prod_{k=1}^{n_\a}(k!)\prod_{1\le j<k\le N}(z_{P(j)}-z_{P(k)}+i\text{sgn}
(y_j-y_k)).
\end{align}
The cluster $\Omega_\a(P)$ is defined by
$\Omega_\a(P)=\{a|\a(P(a))=\b\}$ 
by use of the cluster counting function 
$\a:[1,\cdots,N]\rightarrow[1,\cdots,M] $
such that 
$\a(a)=\b,~{\text{for~}}\sum_{j=1}^{\b-1}n_j<a\le\sum_{j=1}^{\b}n_j$.
The prime means 
the summation over permutations which keep the order inside each cluster, i.e.,
the permutations satisfying the condition 
$j<k \rightarrow P(j)<P(k)$ for $j$ and $k$ 
such that $\a(P(j))=\a(P(k))$ ($1\le j,k\le N$). 

Thus~\eqref{sumeigen} is expressed as
\begin{align}
\sum_{l_1,\cdots,l_N=1}^n
\langle x_{l_1},\cdots,x_{l_N}|\Psi_z\rangle\prod_{j=1}^Ne^{g_{l_j}}
=\sum_{l_1,\cdots,l_N=1}^n{\sum_{P\in S_N}}\!\!{}^{^{'}}
A_P(x_{l_1},\cdots,x_{l_N})e^{\phi(\{l\}_N,P)},
\label{difficulty}
\end{align}
where $\phi(\{l\}_N,P)$ represents the phase
\begin{align}
\phi(\{l\}_N,P)=
i\sum_{\a=1}^Mq_{\a}
\sum_{c\in\Omega_\a(P)}y_c-\frac{1}{4}\sum_{\a=1}^M\sum_{c,c'\in\Omega_\a(P)}
|y_c-y_{c'}|+\sum_{j=1}^Ng_{l_j}
\end{align}
The difficulty in~\eqref{difficulty} is that the summations 
over $l_1,\cdots,l_N$ and $P$ are coupled.
In~\cite{PS2010p,PS2010p2}, the authors proposed an approximation
that the two summations factorize,
\begin{align}
\sum_{l_1,\cdots,l_N=1}^n\sum_{p\in S_N}\!\!{}^{^{'}}
A_P(y_{l_1},\cdots,y_{l_N})e^{\phi(\{l\}_N,P)}
\sim
\sum_{l_1,\cdots,l_N=1}^n
e^{\phi(\{l\}_N)}
\sum_{p\in S_N}\!\!{}^{^{'}}
A_P(y_{l_1},\cdots,y_{l_N}),
\label{factorize}
\end{align}
where
$\phi(\{l\}_N)=\phi(\{l\}_N,P=(1,2,\cdots,N))$.
Under this assumption, we can perform the two summations 
using the relations
\begin{align}
\sum_{P\in S_N}\!\!{}^{^{'}}
A_P(y_{l_1},\cdots,y_{l_N})=\langle 0|\Phi_z\rangle,
\end{align}
which comes from the identity~\eqref{ci1}, and 
\begin{align}
\sum_{l_1,\cdots,l_N=1}^n
e^{\phi(\{l\}_N)}=\prod_{\a=1}^M
e^{-\frac{1}{4}\sum_{j,k=1}^n|x_j-x_k|\partial_{g_j}
\partial_{g_k}}
\left(\sum_{l=1}^ne^{g_l+iq_{\a}x_l}\right)^{n_{\a}},
\label{factorize2}
\end{align}
which is obtained as~(3.13) in~\cite{PS2010p2}.
From~\eqref{factorize}--\eqref{factorize2} and setting 
$y_{j}=2\gamma_t^2X_j$, $g_j=s_j-X_j^2$,
we obtain the approximated form of $B_z$~\eqref{psapprox},
\begin{align}
B_z\sim\prod_{\a=1}^Me^{-\frac12\sum_{j,k=1}^n|X_j-X_k|\partial_{s_j}
\partial_{s_k}}\left(\sum_{l=1}^ne^{-\gamma_ts_l+\gamma_tX_j^2+2i
\gamma_t^2q_{\a}X_{l}}\right)^{n_{\a}}.
\label{appB}
\end{align}
Substituting the equation above into~\eqref{AzBz} 
and following the same procedure in
Sec.4.3, we have 
\begin{align}
G_{\gamma_t}^{\sharp}(\{s\}_n,\{X\}_n)&=1+\sum_{M=1}^{\infty}\frac{1}{M!}
\left(\prod_{\a=1}^M\int_{\R-ic}\frac{dq_\a}
{2\pi}\sum_{n_{\a}=1}^{\infty}\right)
\det\left(\frac{1}{\frac12 (n_\a+n_\b)+i(q_\a-q_\b)}\right)\notag\\
&\hspace{1cm}\times\prod_{\a=1}^M\prod_{r=1}^{n_\a}
\left(\frac{1}{iq_\a+\frac12(n_\a-2r)}\right)\notag\\
&\hspace{1cm}\times\prod_{\a=1}^Me^{\gamma_t^3n_\a^3/12-
\frac12\sum_{j,k=1}^n|X_j-X_k|\partial_{s_j}
\partial_{s_k}}\left(-e^{\frac{tq_\a^2}{2}}\sum_{l=1}^ne^{-\gamma_ts_l+\gamma_tX_j^2+2i
\gamma_t^2q_{\a}X_{l}}\right)^{n_{\a}}.
\end{align}
Here the index $\sharp$ indicates that we use 
the approximation~\eqref{appB}. We find that the equation above has 
a similar structure to the corresponding one in the narrow wedge case
in~\cite{PS2010p2}. All the differences are just two points: 
The contour of $q$ includes the imaginary term $-ic$ and the factor
$\prod_{\a=1}^M\prod_{r=1}^{n_\a}
\left(\frac{1}{iq_\a+\frac12(n_\a-2r)}\right)$ is added. Thus 
we can just follow Sec.~4 in~\cite{PS2010p2} if we replace Eqs.~(4.14) and 
(4.19) in~\cite{PS2010p2} by (1) and (2) in Lemma 6 respectively. 
As a result, we find the Fredholm determinant expression of 
$G^{\sharp}_{\gamma_t}(\{s\}_n,\{X\}_n)$:
\begin{align}
G^{\sharp}_{\gamma_t}(\{s\}_n,\{X\}_n)=
\det(1-L),
\end{align}
where 
\begin{align}
L(z,z')&=\chi_0(z)\chi_0(z')
\exp\left(-\frac12\sum_{j,k=1}^n|X_j-X_k|
\partial_{s_j}\partial_{s_k}
-\gamma_t\sum_{j=1}^nX_j^2\partial_{s_j}
-(\partial_z-\partial_{z'})\sum_{j=1}^nX_j\partial_{s_j}
\right)\notag\\
&\hspace{3cm}
\times\int_{-\infty}^{\infty}du \Ai^{\Gamma}
\left(u+z,\frac{1}{\gamma_t},0\right)
\Ai_{\Gamma}
\left(u+z',\frac{1}{\gamma_t},0\right)
\Phi(\{u-s\}_n).
\label{detL}
\end{align}
Here $\chi_s(x)$ is given below~\eqref{divQ},
$\Ai^\Gamma(a,b,c)$ and $\Ai_\Gamma(a,b,c)$ are defined by~\eqref{AiG1} 
and~\eqref{AiG2} respectively and for
$\Phi(\{x\}_n)$, see~\eqref{Phix}. This expression of the kernel
corresponds to Eq.~(4.25) in~\cite{PS2010p2} for the narrow wedge 
initial condition. The only difference is that
the product of the ordinary Airy functions $\Ai(u+z)\Ai(u+z')$ is
replaced by $\Ai^{\Gamma}(u+z,1/\gamma_t,0)\Ai_{\Gamma}(u+z',1/\gamma_t,0)$
in~\eqref{detL}.

We can further deform the expression~\eqref{detL} following the discussion in
Sec.~4.2 in~\cite{PS2010p2}. The equation corresponding 
to Eq.~(4.38) in~\cite{PS2010p2} becomes
\begin{align}
L(z,z')=\int_{-\infty}^{\infty}&du_1\cdots du_n\langle
\Ai^\Gamma_{z',1/\gamma_t,-X_0/\gamma_t}|L_{0}
e^{(X_0-X_1) H}|u_1\rangle\langle u_1|e^{(X_1-X_2) H}|u_2\rangle\times\cdots
\notag\\
&\times
\langle u_{n-1}|e^{(X_{n-1}-X_n) H}|u_n\rangle
\langle u_n|
e^{(X_n-X_0) H}L_{0}|\Ai_{\Gamma, z,1/\gamma_t,-X_0/\gamma_t}\rangle
\Phi(\{u-s\}_n),
\end{align}
with $X_0=0$.
Here $H$ is the Airy Hamiltonian given below~\eqref{qexpr}, 
$\Phi(\{x\}_n)$ and $L_0$ are defined by~\eqref{Phix} and ~\eqref{Lj} 
respectively. The states $|\Ai^\Gamma_{z,b,c}\rangle$ and 
$|\Ai_{\Gamma,z,b,c}\rangle$ are defined by the relations 
$\langle u|\Ai^\Gamma_{z,b,c}\rangle=\Ai^{\Gamma}(z+u,b,c)$ and $\langle u|\Ai_{\Gamma,z,b,c}\rangle
=\Ai_{\Gamma}(z+u,b,c)$ respectively. In this kernel, we put
$L=Q_2Q_1$, where
\begin{align}
&Q_2(z',u_1)=\langle\Ai^\Gamma_{z',{1}/{\gamma_t},-{X_0}/{\gamma_t}}|L_{0}
e^{-X_1 H}|u_1\rangle,\\
&Q_1(u_1,z)=\int_{-\infty}^{\infty}
du_2\cdots du_{n}
\langle u_1|e^{(X_1-X_2) H}|u_2\rangle\times\cdots\times
\langle u_{n-1}|e^{(X_{n-1}-X_n) H}|u_n\rangle\notag\\
&\hspace{2cm}\times\langle u_n|
e^{X_n H}L_{0}|\Ai_{\Gamma, z,{1}/{\gamma_t},-{X_0}/{\gamma_t}}\rangle
\Phi(\{u-s\}_n).
\end{align}
Noticing the relation of the Fredholm determinant, $\det(1-Q_2Q_1)=
\det(1-Q_1Q_2)$ and the biorthogonality~\eqref{biorthogonality}
of the deformed Airy function, 
we get $G^{\sharp}_{\gamma_t}(\{s\}_n,\{X\}_n)=\det(1-Q)$
where
\begin{align}
&Q(x,y):=Q_1Q_2(x,y)\notag\\
&=\int_{-\infty}^{\infty}du_2\cdots du_n
\langle u_1|e^{(X_1-X_2)H}
|u_2\rangle\cdots \langle u_n|e^{X_nH}L_0^2e^{-X_1H}|y
\rangle \Phi(\{u-s\}_n).
\end{align}
The factor
$\langle x|e^{(X_j-X_k)H}|y\rangle$ is represented 
in terms of the deformed Airy functions as~\eqref{airyairygg},
which will be shown in Appendix.
Using~\eqref{biorthogonality} and (ii) in the Appendix, we find
$L_0^2=L_0$ and $L_0e^{-X_1H}=e^{-X_1H}L_1$, and thus we finally
obtain~\eqref{qexpr}. 
The second last factor $\langle x|e^{tH}L_j|y\rangle$ in~\eqref{qexpr} 
is written as
\begin{align}
\langle x|e^{tH}L_j|y\rangle
=
\int_{0}^{\infty}dze^{-tz}\Ai_{\Gamma}\left(x+z,
\frac{1}{\gamma_t},\frac{-X_j+t}{\gamma_t}\right)
\Ai^{\Gamma}\left(y+z,
\frac{1}{\gamma_t},\frac{-X_j}{\gamma_t}
\right).
\label{0infinity}
\end{align}
This readily follows for $t<0$ from~\eqref{tAiG2} but  
one has to be careful when $t>0$. The integrand
is divergent as $z\rightarrow -\infty$ and hence
$\langle x|e^{tH}|y\rangle$ in~\eqref{airyairygg} is not well-defined.
But in $\langle x|e^{tH}L_j|y\rangle$, $L_j$ projects
the range of integration 
to the positive side only and~\eqref{0infinity}
is still valid.

\subsection{Long-time limit}
Here we consider the long-time limit of the generating function
$G_{\gamma_t}^{\sharp}(\{s\}_n,\{X\}_n)$~\eqref{qexpr}. 
We first introduce the limiting version of $L_j(x,y)$~\eqref{Lj},
\begin{align}
\L_j(x,y)&:=\lim_{\gamma_t\rightarrow\infty}L_j(x,y)=\int_{0}^{\infty}dw \Ai^{-}(x+w,X_j)\Ai^{+}(y+w,X_j),
\end{align}
where $\Ai^{\mp}(x+w,X_j)$ is the limit of the Gamma-deformed Airy 
functions~\eqref{AiG1} and~\eqref{AiG2},
\begin{align}
\Ai^-(x,y)&:=\lim_{\gamma_t\rightarrow\infty}\frac{1}{\gamma_t}\Ai^\Gamma
\left(x,\frac{1}{\gamma_t},\frac{-y}{\gamma_t}\right)
=\frac{1}{2\pi}\int_{\Gamma_{-iy}} dz \frac{e^{izx+\frac{iz^3}{3}}}{iz-y}
\label{Ai-}\\
\Ai^+(x,y)&:=\lim_{\gamma_t\rightarrow\infty}\gamma_t\Ai_\Gamma
\left(x,\frac{1}{\gamma_t},\frac{-y}{\gamma_t}\right)
=\frac{1}{2\pi}\int_{-\infty}^{\infty} dz e^{izx+\frac{iz^3}{3}}(-iz-y).
\label{Ai+}
\end{align}
Here the contour $\Gamma_{-iy}$ in~\eqref{Ai-} 
is given below~\eqref{AiG2}. 
Noticing 
\begin{align}
\lim_{t\rightarrow\infty}\Phi(\{u-s\}_n)=1-\prod_{j=1}^n
\left(1-\Theta(s_j-u_j)\right),
\end{align}
where $\Theta(x)=1~(x\ge 0),~0~(x<0)$, we find the long-time limit
of~\eqref{qexpr} becomes
\begin{align}
\lim_{t\rightarrow\infty}G_{\gamma_t}^{\sharp}(\{s\}_n,\{X\}_n)
=\det\left(1-\L_1+\bar{P}_{s_1}e^{(X_1-X_2)H}\cdots\bar{P}_{s_{n-1}}
e^{(X_{n-1}-X_n)H}
\bar{P}_{s_{n}}e^{(X_n-X_1)H}\L_1 \right).
\label{Gshlimit}
\end{align}
Here $\bar{P}_s$ represents projection operator onto $(-\infty, s)$.
To this equation, we apply the result of Appendix in~\cite{PS2010p2} 
with $\ss{K}_{ij}^0=\delta_{ij}K$ replaced by $\ss{K}_{ij}^0=\delta_{ij}\L_j$.
We eventually obtain the Fredholm determinant with matrix kernel $M$.
\begin{align}
\lim_{t\rightarrow\infty}G_{\gamma_t}^{\sharp}(\{s\}_n,\{X\}_n)
=
\det\left(1-M\right),
\end{align}
where
\begin{align}
M_{jk}(\xi_1,\xi_2)=
\begin{cases}
\chi_{s_j}(\xi_1)\langle\xi_1|e^{(X_j-X_k)H}\L_k|\xi_k\rangle
\chi_{s_k}(\xi_2),& \text{for~} j\ge k,\\
-\chi_{s_j}(\xi_1)\langle\xi_1|e^{(X_j-X_k)H}(1-\L_k)|\xi_2\rangle
\chi_{s_k}(\xi_2),& \text{for~} j<k.\\
\end{cases}
\label{mkernel}
\end{align}
Here $\chi_s(x)$ is given in~\eqref{ltlsh} below.

We find that this matrix kernel is equivalent to 
$\K_{12}(X_j,\xi_1;X_k,\xi_2)$. 
By taking the limit $\gamma_t\rightarrow\infty$ 
in~\eqref{0infinity},
we easily find the factor
$\langle\xi_1|e^{(X_j-X_k)H}\L_k|\xi_k\rangle$ in the case
$j\ge k$ of~\eqref{mkernel} is expressed as
\begin{align}
&~\langle\xi_1|e^{(X_j-X_k)H}\L_k|\xi_2\rangle
\notag\\
&=
\int_{0}^{\infty}dze^{-(X_j-X_k)z}\Ai^-(\xi_1+z,X_j)\Ai^{+}(\xi_2+z,X_k)
\notag\\
&=\int_{-\infty}^{\infty}dw_1\int_{-\infty}^{\infty}dw_2
e^{iw_1\xi_1+iw_2\xi_2+\frac{i}{3}(w_1^3+w_2^3)}
\frac{iw_2+X_k}{(iw_1-X_j)(iw_1+iw_2-X_j+X_k)}.
\end{align}
Here we used~\eqref{Ai-} and~\eqref{Ai+} in the second equality.
Noticing 
\begin{align}
\frac{iw_2+X_k}{(iw_1-X_j)(iw_1+iw_2-X_j+X_k)}=
\frac{1}{iw_1-X_j}-\frac{1}{iw_1+iw_2-X_j+X_k},
\end{align}
we finally find the factor can be represented as the Airy function.
For $j\ge k$, we obtain 
\begin{align}
\langle\xi_1|e^{(X_j-X_k)H}\L_k|\xi_2\rangle&=
\int_{0}^{\infty}dwe^{-w(X_j-X_k)}\Ai(\xi_1+w)\Ai(\xi_2+w)\notag\\
&~+
\Ai(\xi_2)\left(e^{-X_j^3+X_j\xi_1}-\int_{0}^{\infty}dwe^{-X_jw}\Ai(\xi_1+w)\right).
\end{align}
Similarly for the case $j<k$ in~\eqref{mkernel}, we get
\begin{align}
\langle\xi_1|e^{(X_j-X_k)H}\L_k|\xi_2\rangle
&=
-\int_{-\infty}^{0}dwe^{-w(X_j-X_k)}\Ai(\xi_1+w)\Ai(\xi_2+w)\notag\\
&~+
\Ai(\xi_2)\left(e^{-X_j^3+X_j\xi_1}-\int_{0}^{\infty}dwe^{-X_jw}\Ai(\xi_1+w)\right).
\end{align}
Hence we finally find $M_{jk}(\xi_1,\xi_2)=
\K_{12}(X_j,\xi_1;X_k,\xi_2)$.

\section{Directed polymer interpretation}
Our result can also be rephrased for the free energy distribution of the 
1+1 dimensional directed polymer in random media. 
First let us rewrite the path integral expression of 
$Z_{\nu,\lambda,D}(x,t)$~\eqref{path} as 
\begin{align}
Z_{\beta,\gamma,D}(x,t)=\int_{0}^{\infty}dy \int_{x(0)=y}^{x(t)=x}
D[x(\t)]\exp\left[ -\beta (H[x] + \mu B(y)) \right],
\label{Zdp}
\end{align}
where
\begin{align}
H[x]=\int_{0}^{t}d\t \left(\frac{\gamma}{2}\left(\frac{dx}{d\t}\right)^2-
\sqrt{D}\eta(x(\t),\t)\right).
\label{penergy}
\end{align}
The parameters $\beta$, $\gamma$ and $\mu$ are defined as 
\begin{align}
\b=\frac{\lambda}{2\nu},~~\gamma=\frac{1}{\lambda}, ~~
\mu = \frac{\alpha}{\beta}=\sqrt{\frac{D}{2\nu}}
\end{align}
in terms of those of the KPZ equation~\eqref{KPZ}.

This expression (\ref{Zdp}) can be interpreted as the partition function 
of a directed polymer in random media with inverse temperature $\beta$.
On the two dimensional $(x,\t)$~~($x\in\R$ and $\t\ge0$) plane 
the configuration of a polymer is represented  as a function 
$x(\tau)~~(0\le \tau\le t)$. Here it is assumed that $x(\tau)$ is a
single-valued function, which means that the polymer is ``directed''
for $\t$ coordinate. For a configuration $x(\tau)~~(0\le \tau\le t)$
the bulk energy $H[x]$~\eqref{penergy} is assigned. The first 
term of $H[x]$ represents the elastic energy of the polymer whose 
strength is determined by $\gamma$, and the second term represents a 
random potential energy with $D$ adjusting its strength.

The half Brownian motion initial condition~\eqref{stepBM} in the KPZ equation
corresponds to the following boundary condition in the language of the random
directed polymer. At the boundary $\t=0$, the position of the polymer 
$x(0)$ can take any positive value $y\ge 0$ 
and, depending on the position $y$,
the random boundary energy $ \mu B(y)$ is assigned to the 
polymer. Note that the other edge $x(t)$ is pinned at $x$.
As a statistical mechanical model, $\mu$ in (\ref{Zdp}) is an 
independent parameter, but when translated from the KPZ equation 
one has to keep in mind that it is written as 
$\mu= \sqrt{\beta \gamma D}$.

We want to discuss the low temperature behavior of the 
directed polymer for a fixed $t$. 
When the temperature is sufficiently low, the configuration which 
gives the minimum energy is dominant for the free energy. When the position
$x$ of the edge $x(t)$ of the polymer is positive, the other position of 
the edge $y$ tends to be close to $x$ because otherwise the polymer
obtains excess bulk elastic energy. In this case, it is dominated by 
the boundary energy $B(y)$. 
Thus in the positive region $x>0$, the free energy fluctuation 
is described by the Gaussian.
On the other hand, when $x<0$, the edge $x(0)$ tends to be pinned at 
the origin. In this case, the free energy fluctuation is determined 
only by the bulk energy (note that $B(y=0)=0$). In between around $x=0$ 
one observes the crossover distribution, which is exactly what have been 
computed in this paper. 

Concretely, from~\eqref{ColeHopf}, 
$-h_{\nu,\lambda,D}$ corresponds to the free energy 
$f_{\beta,\gamma,D}(x,t)$ of the random directed polymer
and hence (\ref{hdist}) is translated as
\begin{align}
\text{Prob}\left(\beta f_{\beta,\gamma,D}(x,t)
-\frac{\gamma_{t}^3}{12}+\gamma_tX^2
\ge \gamma_{t} s
\right)=F_{\gamma_t}(s;X),
\label{mainresultarbitrary2}
\end{align}
where $F_{\gamma_t}(x;X)$ is given in~\eqref{Th3} 
and $\a$ and $\gamma_t$ are written in terms of $\gamma$ and $\beta$ as
$\a=(\beta^3\gamma D)^{1/2}$ and 
$\gamma_t=(\beta^5\gamma D^2 t/2)^{1/3}$ respectively.
Based on (\ref{mainresultarbitrary2}), 
one can obtain various information about the the free energy of 
the directed polymer. For instance one sees that as $\beta\to\infty $ 
the macroscopic free energy scales as $O(\beta^4)$, the fluctuations of the 
bulk free energy is $O(\beta^{2/3})$. 
To get a nontrivial distribution in the low temperature limit
$\beta\to\infty$, one has to scale the space direction as 
$X = O(\beta^{1/3})$. 
The free energy distribution in this limit is given by~\eqref{Cor4} 
since $\beta\to\infty$ implies $\gamma_t\to\infty$.
This describes the crossover between boundary and bulk free 
energy fluctuation, i.e., between Gaussian and the GUE
Tracy-Widom distribution.

\section{Conclusion}
In this paper we have considered the KPZ equation~\eqref{KPZ} with
the half Brownian motion initial condition~\eqref{stepBM}.
Using the Bethe ansatz results of the one-dimensional attractive 
$\delta$-Bose gas, we have obtained a Fredholm determinant expression for
the generating function of exponential moments of the height (Theorem 1).
Thanks to this result and Proposition 2, the compact representation
of the probability distribution of the height 
was obtained (Theorem 3). We have also found an expression for the
multi-point generating function employing an approximation 
proposed in~\cite{PS2010p,PS2010p2}. 

These results are expressed in terms of the deformed Airy 
functions~\eqref{AiG1} and~\eqref{AiG2}. If we change 
these functions to the ordinary Airy function, 
we recover the
results for the narrow wedge initial
condition obtained in~\cite{SS2010a,SS2010b,SS2010c,ACQ2010}.
The deformed Airy functions have some nice properties, which
are discussed in the Appendix. They satisfy the biorthogonality 
relation. Their time evolution by the Airy Hamiltonian are again
given by the same functions with a parameter modified.
We remark that similar relations of the multiple Hermite polynomials 
played an important role in the study of the PNG model with external 
source \cite{IS2004,IS2005}. 
In the long time limit, the kernel of the Fredholm determinant becomes 
the rank one perturbation
of the Airy kernel. It is not clear whether the higher rank 
perturbations studied in~\cite{BBP2006} have corresponding finite time
generalizations.

\appendix
\section
{Properties of the deformed Airy functions}
In this appendix, we pick up a few properties 
of the gamma-deformed Airy functions~\eqref{AiG1} and~\eqref{AiG2},
which are necessary for the discussions in Sec.~6.
Let $t\le 0$ and $H$ be the Airy Hamiltonian 
$H=-\partial^2/\partial x^2+x$. We have the following relations.

\vspace{3mm}
\noindent

\noindent
(i) The deformed Airy function representation of the propagator
\begin{align}
\langle x|e^{tH}|y\rangle
=\int_{-\infty}^{\infty}dze^{-tz}\Ai_{\Gamma}\left(x+z,
b,c-bt\right)
\Ai^{\Gamma}\left(y+z,b,c
\right).
\label{drprop}
\end{align}
Note that when we set $t=0$ in~\eqref{drprop}, we get the biorthogonality 
relation of the deformed Airy function,
\begin{align}
\int_{-\infty}^{\infty}dw \Ai_{\Gamma}(x+w,b,c)\Ai^{\Gamma}(y+w,b,c)
=\delta(x-y).
\label{biorthogonality}
\end{align}

Using~\eqref{drprop}, we easily get the following relations.

\noindent
(ii) ``Time evolution" by the Airy Hamiltonian
\begin{align}
&e^{tH}\Ai^{\Gamma}({x+w,b,c})=e^{-tw}\Ai^{\Gamma}({x+w,b,c-bt}),
\label{tAiG1}
\\
&e^{tH}\Ai_{\Gamma}({x+w,b,c})=e^{-tw}\Ai_{\Gamma}({x+w,b,c+bt}).
\label{tAiG2}
\end{align}

When we set $b=1/\gamma_t$ and $c=-X/\gamma_t$ and take the limit
$\gamma_t\rightarrow\infty$, these 
relations~\eqref{drprop},~\eqref{tAiG1} and \eqref{tAiG2}
become those for $\Ai^-(x,y)$~\eqref{Ai-} 
and $\Ai^+(x,y)$~\eqref{Ai+},
\begin{align}
&\langle x|e^{tH}|y\rangle
=\int_{-\infty}^{\infty}dze^{-tz}\Ai^+\left(x+z,
X\right)
\Ai^{-}\left(y+z,X
\right),\\
&e^{tH}\Ai^{-}({x+w,X})=e^{-tw}\Ai^{-}({x+w,X-t}),
\label{ltAiG-}
\\
&e^{tH}\Ai^{+}({x+w,X})=e^{-tw}\Ai^{+}({x+w,X+t}).
\label{ltAiG+}
\end{align}

\noindent
{\bf Proof} 

\noindent
(i)
The factor $\langle x|e^{tH}|y\rangle$ can be represented 
in terms of the Airy function,
\begin{align}
\langle x|e^{tH}|y\rangle=\int_{-\infty}^{\infty}dw
e^{-tw}\Ai(w+x)\Ai(w+y).
\label{pAiry}
\end{align}
Thus it is enough to show that
\begin{align}
\int_{-\infty}^{\infty}dw e^{-tw}\Ai_{\Gamma}\left(x+w,
b,c-bt\right)
\Ai^{\Gamma}\left(y+w,b,c
\right)
=
\int_{-\infty}^{\infty}dw
e^{-tw}\Ai(w+x)\Ai(w+y).
\label{irelation}
\end{align}
In the definition~\eqref{AiG1} of $\Ai^{\Gamma}(a,b,c)$, 
we find that in the case $c>0$, 
the contour integral $\Gamma_{ic/b}$ can be 
replaced by $\R$ while in the case $c\leq 0$,
we can divide the contour integral into
the integral on $\R$ and the contributions from the poles at
$(ic+m)/b$, $m=0,1,2\cdots$ of $\Gamma(ibz+c)$. Thus 
\eqref{AiG1} can be represented as
\begin{align}
\Ai^{\Gamma}(a,b,c)=
\begin{cases}
\frac{1}{2\pi}\int_{-\infty}^{\infty} dz 
e^{iza+i\frac{z^3}{3}}{\Gamma\left(ibz+c\right)}, & c>0,\\
\frac{1}{2\pi}\int_{-\infty}^{\infty} dz 
e^{iza+i\frac{z^3}{3}}{\Gamma\left(ibz+c\right)}+\sum_{N=0}^n
\frac{(-1)^N}{N!b}e^{-\frac{(c+N)a}{b}+\frac{(c+N)^3}{3b^3}}, & c \leq 0.
\end{cases}
\label{AiG1r}
\end{align}
Here $n$ is the maximum integer satisfying $c+n<0$.

Thus for the case $c>0$, using~\eqref{AiG2} and the first relation in~\eqref{AiG1r}, we find the left hand side of~\eqref{irelation} is represented as
\begin{align}
&~\frac{1}{2\pi}\int_{-\infty}^{\infty}dw
\int_{-\infty}^{\infty}dz_1
\int_{-\infty}^{\infty}dz_2
e^{iw(z_1+z_2+it)+i(z_1x+z_2y)+\frac{i}{3}(z_1^3+z_2^3)}
\frac{\Gamma(ibz_2+c)}{\Gamma(-ibz_1+c-bt)}\notag\\
&=
~\frac{1}{2\pi}\int_{-\infty}^{\infty}dw
\int_{\R-it}dz_1
\int_{-\infty}^{\infty}dz_2
e^{iw(z_1+z_2+it)+i(z_1x+z_2y)+\frac{i}{3}(z_1^3+z_2^3)}
\frac{\Gamma(ibz_2+c)}{\Gamma(-ibz_1+c-bt)}.
\end{align}
Here we deformed the contour of $z_1$ to $\R-it$. This
deformation is valid when $t\le 0$. Note that in this equation above, 
the part $z_1+z_2+it$ becomes real. Thus 
one has
\begin{align}
&~\int_{\R-it}dz_1
\int_{-\infty}^{\infty}dz_2
\delta(z_1+z_2+it)
e^{i(z_1x+z_2y)+\frac{i}{3}(z_1^3+z_2^3)}
\frac{\Gamma(ibz_2+c)}{\Gamma(-ibz_1+c-bt)}\notag\\
&=
\int_{\R-it}dz_1
\int_{-\infty}^{\infty}dz_2
\delta(z_1+z_2+it)
e^{i(z_1x+z_2y)+\frac{i}{3}(z_1^3+z_2^3)}\notag\\
&=
\frac{1}{2\pi}\int_{-\infty}^{\infty}dw
\int_{-\infty}^{\infty}dz_1
\int_{-\infty}^{\infty}dz_2
e^{iw(z_1+z_2+it)+i(z_1x+z_2y)+\frac{i}{3}(z_1^3+z_2^3)}.
\label{Ajkxy2}
\end{align}
In the second equality, the factor of the Gamma functions is eliminated because
of the delta function. We easily find that the last expression of this equation
is equal to the right hand side of~\eqref{irelation} from the 
integral representation of 
the Airy function,
\begin{align}
\Ai(x)=\int_{-\infty}^{\infty}dz e^{izx+i\frac{z^3}{3}}.
\end{align}

For $c<0$, on the other hand, the terms coming from the pole contributions
in~\eqref{AiG1r} are added. However we find that these terms do not affect
the result since
\begin{align}
&~\int_{-\infty}^{\infty}dwe^{-tw}\Ai_{\Gamma}(x+w,b,c)
e^{-\frac{(c+N)}{b}(y+w)}
\notag\\
&=
\frac{1}{2\pi}\int_{-\infty}^{\infty}dw\int_{-\infty}^{\infty}
dz_1\frac{e^{i\left(z_1+i\frac{c+N}{b}+it\right)w-\frac{c+N}{b}y+ix z_1+i
\frac{z_1^3}{3}}}{\Gamma(-ibz_1+c-bt)}
\notag\\
&=
\int_{\R-i\left(\frac{c+N}{b}+t\right)}dz_1
\delta\left(z_1+i\frac{c+N}{b}+it\right)
\frac{e^{-\frac{c+N}{b}y+iz_1x}}{\Gamma(-ibz_1+c-bt)}
=
\frac{e^{\frac{c+N}{b}(x-y)+tx}}{\Gamma(-N)}
=0.
\end{align}

\noindent
(ii) Here we only consider~\eqref{tAiG1} and omit the proof of~\eqref{tAiG2}
since it can be shown by using the same strategy as~\eqref{tAiG1}.
The left hand side is given by
\begin{align}
&~\langle x|e^{tH}|\Ai^{\Gamma}_{w,b,c}\rangle
=\int_{-\infty}^{\infty}dy\langle x|e^{tH}|y\rangle 
\Ai^{\Gamma}(y+w,b,c).
\label{timeG2}
\end{align}
Substituting~\eqref{drprop} into this equation, we readily obtain
the right hand side of~\eqref{tAiG1},
\begin{align}
\langle x|e^{tH}|\Ai^{\Gamma}_{w,b,c}\rangle
&=\int_{-\infty}^{\infty}dz\int_{-\infty}^{\infty}dye^{-tz}
\Ai^{\Gamma}(x+z,b,c-bt)\Ai_{\Gamma}(y+z,b,c)
\Ai^{\Gamma}(y+w,b,c)\notag\\
&=e^{-tw}
\Ai^{\Gamma}(x+w,b,c-bt).
\end{align}
Here in the second equality, we used the biorthogonality 
relation~\eqref{biorthogonality}.
\begin{flushright}
$\blacksquare$
\end{flushright} 

\section*{Acknowledgments}
The work of T.I. and T.S. is supported by 
KAKENHI (22740251) and KAKENHI (22740054) respectively.

\end{document}